\documentclass[12pt]{article}
\usepackage{amsmath}
\usepackage{graphicx}
\usepackage{enumerate}
\usepackage{natbib}
\usepackage{url} 
\usepackage{hyperref}
\usepackage{gensymb}

\newcommand{\blind}{1}

\usepackage{chapterbib}
\usepackage{multirow}

\newcommand{\cov}{\hbox{cov}}
\newcommand{\trans}{^{ \mathrm{\scriptscriptstyle T} }}
\newcommand{\inv}{^{-1}}
\newcommand{\vecop}{\hbox{vec}}
\newcommand{\bs}{\mathbf{s}}

\newcommand{\bw}{\boldsymbol{w}}
\newcommand{\by}{\boldsymbol{y}}
\newcommand{\bY}{\mathbf{Y}}
\newcommand{\bx}{\boldsymbol{x}}
\newcommand{\bX}{\mathbf{X}}
\newcommand{\bW}{\mathbf{W}}
\newcommand{\btheta}{\boldsymbol{\theta}}

\newcommand{\blambda}{\boldsymbol{\lambda}}
\newcommand{\bxi}{\boldsymbol{\xi}}

\newcommand{\bchi}{\boldsymbol{\chi}}
\newcommand{\bmu}{\boldsymbol{\mu}}

\newcommand{\bone}{\boldsymbol{1}}

\begin{document}
	
	\def\spacingset#1{\renewcommand{\baselinestretch}%
		{#1}\small\normalsize} \spacingset{1}

	\if1\blind
	{
		\title{\bf Combining interdependent climate model outputs in CMIP5: A spatial Bayesian approach}
		\author{Huang Huang \thanks{huang.huang@kaust.edu.sa}\hspace{.2cm}\\
			CEMSE Division,\\
			King Abdullah University of Science and Technology\\
			and \\
			Dorit Hammerling \\
			Department of Applied Mathematics and Statistics,\\
			Colorado School of Mines\\
			and \\
			Bo Li\\
			Department of Statistics,\\
			University of Illinois at Urbana-Champaign\\
			and\\
			Richard Smith\\
			Department of Statistics and Operations Research, \\
			University of North Carolina, Chapel Hill\\
			\\
			\\
		}
		\maketitle
	} \fi
	
	\if0\blind
	{
		\bigskip
		\bigskip
		\bigskip
		\begin{center}
			{\LARGE\bf Combining interdependent climate model outputs in CMIP5: A spatial Bayesian approach}
		\end{center}
		\medskip
	} \fi
	
	\bigskip

	\begin{abstract}
	Projections of future climate change rely heavily on climate models, and combining climate models through a multi-model ensemble is both more accurate than a single climate model and valuable for uncertainty quantification. However, Bayesian approaches to multi-model ensembles have been criticized for making oversimplified assumptions about bias and variability, as well as treating different models as statistically independent. This paper extends the Bayesian hierarchical approach of \citet{Sansom2017} by explicitly accounting for spatial variability and inter-model dependence. We propose a Bayesian hierarchical model that accounts for bias between climate models and observations, spatial and inter-model dependence, the emergent relationship between historical and future periods, and natural variability. Extensive simulations show that our model provides better estimates and uncertainty quantification than the commonly used simple model mean. These results are illustrated using data from the CMIP5 model archive. As examples, for Central North America our projected mean temperature for 2070--2100 is about 0.8 K lower than the simple model mean, while for East Asia it is about 0.5 K higher; however, in both cases, the widths of the $90\%$ credible intervals are of the order 3--6 K, so the uncertainties overwhelm the relatively small differences in projected mean temperatures.

	\end{abstract}

	\noindent%
	\vfill
	
	\newpage
	\spacingset{1.5} 
	
	\addtolength{\textheight}{.5in}%
	
	\section{Introduction}\label{sec:intro}

The reports of the Intergovernmental Panel on Climate Change (IPCC) provide regular updates on the state of climate science; the most recent report was the Fifth Assessment Report published in 2013~\citep{Stocker2013}; the Sixth Assessment Report is due to be published
in 2021. The science in these reports relies heavily on climate models, which form the basis for projections of future climate under a variety of assumptions about greenhouse gases and other anthropogenic emissions. The Coupled Models Intercomparison Project, version 5, popularly known as CMIP5, is a compilation of climate model data from modeling groups around the world~\citep{Taylor2012}. These models allow the user to calculate projections for a very large 
number of meteorological variables, on a wide variety of spatial and temporal scales. For the Sixth Assessment Report, CMIP5 will be replaced by a considerably expanded set of model simulations, CMIP6, but results from these simulations are not yet generally available.

From an early stage of the development of climate science around large modeling exercises of this nature, it has been generally recognized that there are many advantages to be gained by combining results from different climate models rather than by treating the models one at a 
time --- known as the \textit{multi-model ensemble} approach. \citet{Raisanen2001} pioneered an explicit probabilistic approach and assumed equal weighting over all the models. This was quickly contrasted, however, by the \textit{Reliability Ensemble Average} approach~\citep{Giorgi2003}, which weighted models according to their agreement with historical data as well as taking account of how well future projections from different models
agreed with each other.

This led to a series of papers taking a Bayesian statistics approach in which prior distributions were placed on certain unknown model parameters and a posterior predictive approach was taken to derive probabilistic projections for future climate variables~\citep{Tebaldi2004,Tebaldi2005,Min2006,Tebaldi2009,Smith2009}. However, as first noted by \citet{Greene2006} and elaborated further by \citet{Tebaldi2007}, these simple Bayesian approaches may not produce realistic projections of uncertainty. They essentially treated climate model projections as independent perturbations of some unknown ``true" climate variable, ignoring both systematic biases between models and observations and the fact that many climate models tend to be correlated. The latter statement is true, in part, because of direct collaboration between modeling groups (for example, many of the supposedly different models in CMIP5 are actually different versions of climate models produced by the same modeling group), but even in the absence of such collaboration, climate models from different modeling groups use similar physical assumptions and computational methodology, so it is natural to expect that their errors will be correlated.  \citet{BoLi2016} also indicated that some climate models are in a closer agreement than others.

Another comment about these early approaches to multi-model ensembles is that most statistical approaches were either for the marginal distribution of a single climate variable or, at most, the joint distribution of a small number of climate variables, e.g., temperature
and precipitation~\citep{Tebaldi2009}. However, \citet{Furrer2007} made an early attempt at extending the approach to a spatially-correlated random field.

Over the past decade, these approaches have been greatly extended to allow for more complex and realistic representations of model error and the associated uncertainties. \citet{Buser2009} allowed for model bias (the systemic discrepancies between model output and observations that are not eliminated by repeated sampling) and also
considered the effect of inter-annual variability.

\citet{Chandler2013} summarized the strength and weaknesses of Bayesian approaches to multi-model ensembles, noting, for example, that such approaches could fail if there were errors common to all the models, and also that approaches such as the
Reliability Ensemble Average, although designed to give higher weight to the more reliable models, in practice often performed worse than simple uniform averaging over all the models~\citep{Weigel2010}. He proposed an alternative approach, independently developed by \citet{Rougier2013}, which was based on the notion of exchangeability and characterized by \citet{Chandler2013} as ``reality is treated essentially as though it were another simulator." In this paradigm, increasing the number of simulators will not necessarily reduce the uncertainty to zero. However, the approach in effect decomposes
the errors in a climate model as the sum of its deviation from some overall average ``consensus" model, and the deviation between the consensus model and the true Earth system.

Another idea to appear around the same time was that of an ``emergent relationships" (or ``emergent constraints"), which refers to some fixed relationships that are common to all climate models; \citet{Bracegirdle2012,Bracegirdle2013} showed examples for Arctic sea ice. In effect, such a model would imply the existence of some fixed parameter(s) describing  the relationships between present and future climates 
that are common to all models.

These ideas have all been brought together in the recent paper~\citep{Sansom2017} which, to quote the authors, ``accounts for model uncertainty, model inadequacy, internal variability, natural variability, observation uncertainty and emergent relationships". However, it does not account for spatial correlation nor dependence between climate models.

The model dependence issues have also been addressed in previous literature. \citet{Bishop2013} and \citet{Abramowitz2015} pointed out this possible climate model dependence issue and introduced a \textit{replicate Earth paradigm} to seek model dependence from error correlation in some transformed ensemble projections. A recent paper by \citet{Abramowitz2019}  has reviewed all the current approaches to model dependence and discussed their possible application to the forthcoming CMIP6 ensemble.

The present paper extends the model proposed by \citet{Sansom2017} by incorporating spatial correlation and dependence between climate models. In common with the main graphical model by \citet{Sansom2017}, we propose in Section 2 a hierarchical model whose components include climate model outputs and observations, latent variables, and model parameters, but the main objects considered are spatial random fields, represented by Gaussian processes with parametric covariance functions. We propose an MCMC sampling approach to estimate the parameters of the spatial random field, a parameter representing
an emergent relationship, and also, a covariance matrix for inter-model dependence. The latter therefore allows, explicitly, for the possibility that different climate models may be dependent because of common modeling strategies or for other reasons that may cause climate models by different modeling groups to produce similar results. Our extension to take into account various forms of dependence results in more precise uncertainty quantification.

The rest of the paper is organized as follows. Section~\ref{sec:model} describes the hierarchical model and MCMC sampling strategy in detail. Section~\ref{sec:simulation} contains detailed simulations to understand how the model performs on simulated synthetic data. Section~\ref{sec:application} then shows how the method applies to real-data
examples, the near-surface temperature in Central North America and East Asia. Section~\ref{sec:discussion} summarizes the benefits of using our approach in analyzing CMIP5 model outputs and points out directions for potential improvement that may be used in processing upcoming CMIP6 results.
	
	\section{Bayesian hierarchical model}\label{sec:model}
A typical climate model generates averages of a meteorological variable, such as temperature or precipitation over a finite set of grid cells at a specified temporal resolution. In our modeling framework, we assume that the data we obtained from the climate models are integrated over time to create a map of historical and future means.

\subsection{Climate model layers}\label{subsec:climateModel}
Climate models are executed under specific initial conditions and generate realizations that bring internal variability, where each realization is called a climate model run.
We assume that the realizations of each climate model have a different underlying model mean deviating from the consensus field that all climate models agree.
Let $\mathcal{D}$ be the spatial grid for the study domain. For a particular climate variable of interest, we denote the ensemble consensus random field of all the climate models in the historical and future periods at location $\bs\in\mathcal{D}$ by $\mu_H(\bs)$ and $\mu_F(\bs)$, respectively.

Suppose that there is a total number of $M$ available climate models. For each model $m=1,\ldots,M$, the underlying climate model mean in the historical and future periods at location $\bs$ are denoted by $X_{Hm}(\bs)$ and $X_{Fm}(\bs)$, respectively. Considering that the historical and future periods are set far apart, they are assumed to have independent noises. Then, the proposed statistical models for $X_{Hm}(\bs)$ and $X_{Fm}(\bs)$ are as follows,
\[
\begin{array}{rcl}
X_{Hm}(\bs)&=&\mu_H(\bs)+\epsilon_{Hm}(\bs),\\
X_{Fm}(\bs)&=&\mu_F(\bs)+\epsilon_{Fm}(\bs)+\beta\{X_{Hm}(\bs)-\mu_H(\bs)\},
\end{array}
\]
where the remainder random processes $\epsilon_{Hm}(\bs)$ and $\epsilon_{Fm}(\bs)$ are the two independent noise processes, assumed to be zero-mean spatial Gaussian processes. The additional term in the climate model mean in the future period accounts for the consistent bias in the climate model means in the two periods, and the coefficient $\beta$ is called the emergent relationship. We assume the covariance functions for the two processes  $\epsilon_{Hm}(\bs)$ and $\epsilon_{Fm}(\bs)$ are as follows:
\begin{equation}\label{eq:covForClimateModels}
\begin{array}{rcl}
\cov\big(\epsilon_{Hp}(\bs_i),\epsilon_{Hq}(\bs_j)\big)&=&{\tau\inv_H}c(\|\bs_i-\bs_j\|;\gamma_H)v_{pq},\\
\cov\big(\epsilon_{Fp}(\bs_i),\epsilon_{Fq}(\bs_j)\big)&=&{\tau\inv_F}c(\|\bs_i-\bs_j\|;\gamma_F)v_{pq},\\
\end{array}
\end{equation}
where $\bs_i$, $\bs_j$ are two arbitrary locations in the spatial domain $\mathcal{D}$, and $p, q = 1, \ldots, M$ are any two climate model indices, $\tau_H$ and $\tau_F$ are the inverse of sill parameters accounting for the spatial variance, $\gamma_H$ and $\gamma_F$ are the range parameters in the Whittle covariance function $c(\cdot;\gamma_H)$, $c(\cdot;\gamma_F)$, and $v_{pq}$ controls the non-spatial correlation between model $p$ and model $q$. We write all the  $v_{pq}$'s for $p,q=1,\ldots,M$ as a matrix $V$ with $(p,q)$-th entry $v_{pq}$. It is easy to observe that the covariance functions would remain the same if a constant is multiplied to $\tau_H$, $\tau_F$, and $V$ all together. Therefore, to make $\tau_H$ and $\tau_F$ identifiable, $v_{11}$ is fixed to be one, which implies $\tau_H\inv$ and $\tau_F\inv$ are the variances at any locations in the climate model 1. It is noteworthy that ${\tau\inv_H}c(\|\bs_i-\bs_j\|;\gamma_H)$ and ${\tau\inv_F}c(\|\bs_i-\bs_j\|;\gamma_F)$ describe the spatial covariance while $V$ accounts for the climate model dependence. Thus, a separable covariance structure is essentially assumed between these two types of covariances. 

Note that the climate model means $X_{Hm}(\bs)$ and $X_{Fm}(\bs)$ are latent states that we do not observe. The data we have in the climate model outputs are individual climate model runs simulated from the corresponding climate model with a particular initial condition. Therefore, it is natural to treat different climate model runs as the corresponding climate model mean plus some noise associated with internal variabilities. The internal variability may not lead to white noise and the correlation among different locations could exist. Thus, we also use spatial Gaussian processes to model the internal variability. Then, for each climate model $m$, the climate model run $r=1,\ldots,R_{Hm}$ in the historical period and the climate model run $r^\prime=1,\ldots,R_{Fm}$ in the future period, where $R_{Hm}$ and  $R_{Fm}$ are the total number of model runs for model $m$ in the historical and future periods, respectively, are modeled as
\[
\begin{array}{rcl}
X_{Hmr}(\bs)&=& X_{Hm}(\bs)+\epsilon_{Hmr}(\bs),\\
X_{Fmr^\prime}(\bs)&=& X_{Fm}(\bs)+\epsilon_{Fmr^\prime}(\bs),
\end{array}
\]
where $\epsilon_{Hmr}(\bs)$ and $\epsilon_{Fmr^\prime}(\bs)$ are zero-mean Gaussian processes with covariance function
\begin{equation}\label{eq:modelRuns}
\begin{array}{rcl}
\cov\big(\epsilon_{Hmr}(\bs_i),\epsilon_{Hmr}(\bs_j)\big)&=&\phi_{Hm}\inv c(\|\bs_i-\bs_j\|;\gamma_{Hm}),\\ \cov\big(\epsilon_{Fmr^\prime}(\bs_i),\epsilon_{Fmr^\prime}(\bs_j)\big)&=&\phi_{Fm}\inv c(\|\bs_i-\bs_j\|;\gamma_{Fm}).
\end{array}
\end{equation}
The inverse of the sill parameters $\phi_{Hm}$ and $\phi_{Fm}$ are assumed to follow the conjugate distributions, which are Gamma distributions as follows,
\begin{equation}\label{eq:phiHmFm}
\phi_{Hm}\sim Ga(\frac{\nu_H}{2},\frac{\nu_H\phi_H\inv}{2}), ~\phi_{Fm}\sim Ga(\frac{\nu_F}{2},\frac{\nu_F\phi_F\inv}{2}),
\end{equation}
where $\nu_H$, $\phi_H$, $\nu_F$, and $\phi_F$ are unknown hyper-parameters.
\subsection{Observation Layers}
Section~\ref{subsec:climateModel} describes how the climate model output is related to the underlying ensemble consensus field. On the other hand, since all the climate models considered in this work try to mimic the climate of the real world, the consensus field is also linked to the real-world climate and subsequently the observations. We elaborate these connections in this section. We call the climate that is actually occurring on the earth the actual climate and treat it as a random realization from a particular distribution, the mean of which is called the expected climate and denoted by $Y_H(\bs)$ and $Y_F(\bs)$, for the historical and the future periods, respectively. Since the climate models attempt to simulate the real-world climate, the expected climate can also be treated as a realization from the ensemble consensus field with some uncertainty, which was also used and discussed by \citet{Sansom2017}. That being said, the expected climate is viewed as a counterpart to the climate model mean described in Section~\ref{subsec:climateModel}. Thus, the same statistical models as used for the climate model means $X_{Hm}(\bs)$ and $X_{Fm}(\bs)$ are assumed for the expected climate. More specifically,
\begin{equation}\label{eq:expectedClimate}
\begin{array}{rcl}
Y_H(\bs)&=& \mu_H(\bs)+\epsilon_H(\bs),\\
Y_F(\bs)&=& \mu_F(\bs)+\epsilon_F(\bs)+\beta\{Y_{H}(\bs)-\mu_H(\bs)\}.\\
\end{array}
\end{equation}
The remainder processes $\epsilon_{H}(\bs)$ and $\epsilon_{F}(\bs)$ are also assumed to be zero-mean spatial Gaussian processes. However, the covariance function has no model dependence component but a predetermined scaling factor $\kappa > 0$ in front of the spatial covariance accounting for the potential inadequacy of climate models in characterizing the variability, i.e.,
\[
\begin{array}{rcl}
\cov(\epsilon_H(\bs_i),\epsilon_H(\bs_j))&=&\kappa\tau\inv_Hc(\|\bs_i-\bs_j\|;\gamma_H),\\
\cov(\epsilon_F(\bs_i),\epsilon_F(\bs_j))&=&\kappa\tau\inv_Fc(\|\bs_i-\bs_j\|;\gamma_F).\\
\end{array}
\]

Then, the actual climate $Y_{Ha}(\bs)$ and $Y_{Fa}(\bs)$ for the historical and the future periods, respectively, should be the expected climate plus some noise corresponding to the natural variability, and we use white noise to model these natural variabilities as shown below,
\[
\begin{array}{rcl}
Y_{Ha}(\bs)&\sim& N\big(Y_H(\bs),\phi_{Ha}\inv\big),\\
Y_{Fa}(\bs)&\sim& N\big(Y_F(\bs),\phi_{Fa}\inv\big).\\
\end{array}
\]
The precision parameters of the white noise $\phi_{Fa}$ and $\phi_{Ha}$ are assumed to follow Gamma distributions with the same mean as $\phi_{Hm}$ and $\phi_{Fm}$ shown in Formula~\eqref{eq:phiHmFm} but different variances, respectively. More specifically,
$\phi_{Ha}\sim Ga(\nu_H/(2\kappa),\nu_H\phi_H\inv/(2\kappa))$ and $\phi_{Fa}\sim Ga(\nu_F/(2\kappa),\nu_F\phi_F\inv/(2\kappa))$.
We see that the mean of $\phi_{Ha}$ and $\phi_{Hm}$ ($\phi_{Fa}$ and $\phi_{Fm}$) are the same, but the shape and rate parameters in the assumed Gamma distributions are different with a common scaling factor $\kappa$. The parameter $\kappa$ accounts for how the actual climate might be different than a realization from another ``climate model'' due to the potential inadequacy of the internal variability (as expressed through the implemented climate models) to truly capture natural variability.

\begin{figure}[ht!]
	\centering
	\includegraphics[width=\linewidth]{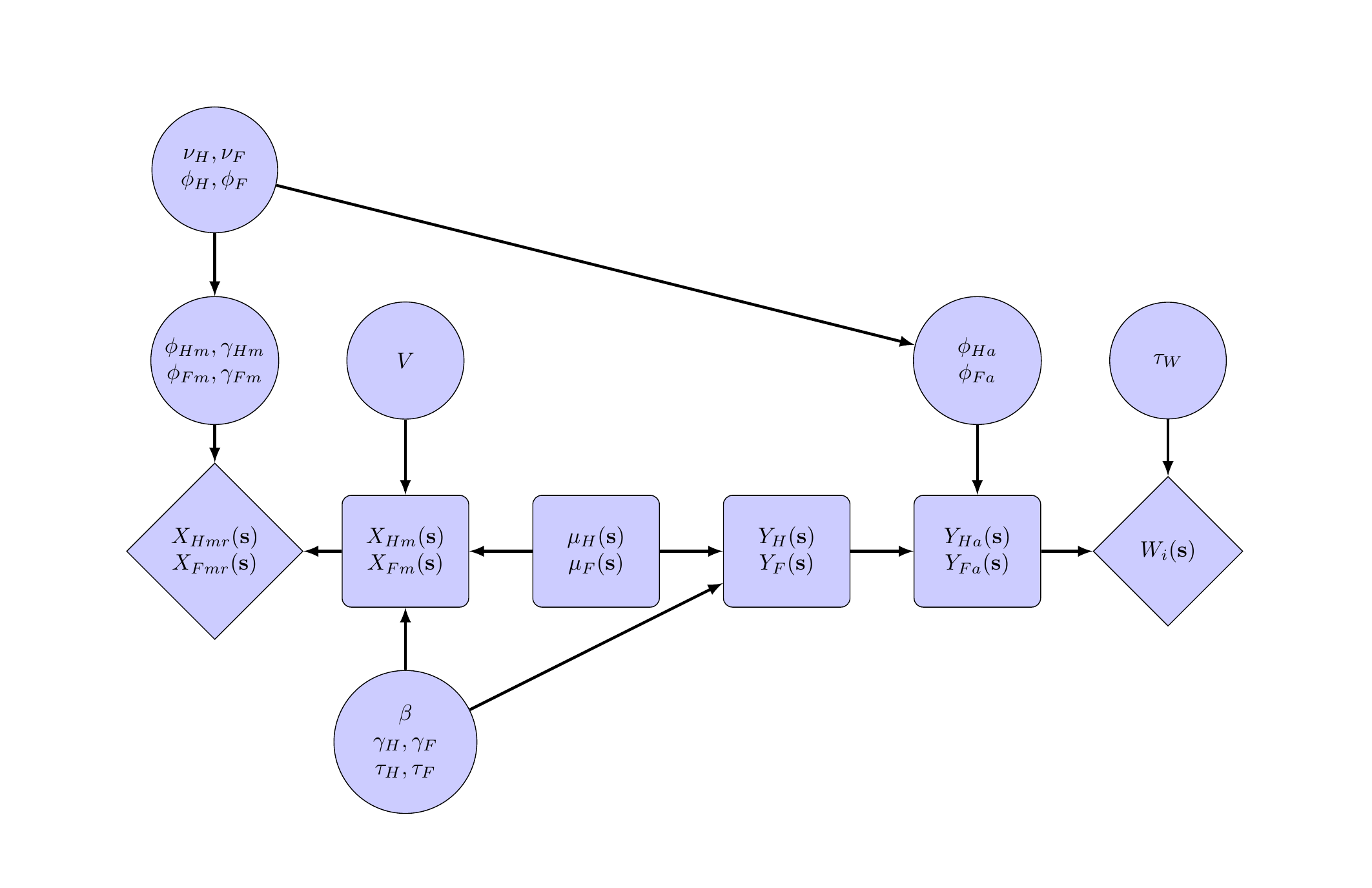}
	\caption{Illustration of the proposed Bayesian hierarchical model.  Diamonds represent the available data, rectangles represent the latent states, and circles represent the parameters.}
	\label{fig:model}
\end{figure}

Finally, because observations can have measurement errors or errors from other sources on top of the actual climate, the observations denoted by $W_i(\bs)$ for $i=1,\ldots,N$, where $N$ is the total number of observational data sets, are modeled as actual climate plus white noise, i.e.,
\[
W_i(\bs)\sim N\big(Y_{Ha}(\bs),\tau_W\inv\big),
\]
where $\tau_W$ is the unknown hyper-parameter for the precision of the white noise. 
We summarize our full Bayesian hierarchical model in Figure~\ref{fig:model}, where the relationship among three types of variables---data, latent states, and parameters---is illustrated. 

\subsection{Inference}
The advantage of using a Bayesian hierarchical model is that it naturally integrates all the model components in different layers into a single framework, and all uncertainties in different layers propagate to the final results. 
However, the obtained posterior distributions of latent states or parameters are not only conditional on the data but also depend on the prior distributions we specified for the parameters. To alleviate the effects of priors, we use non-informative prior distributions whenever possible, hoping that the priors will have less influence on the posterior results.

The prior distributions for all the parameters are specified as follows. We propose normal prior distributions for $\mu_H(\bs)$ and $\mu_F(\bs)$ as $\mu_H(\bs), \mu_F(\bs)\sim N(0,10^{6})$, and an inverse Wishart prior distribution for $V$ as $V\sim IW(d\tilde V,M+d+1)$, where $\tilde{V}$ is the mean of $V$ in the prior and $d$ is a predetermined integer that controls the degrees of freedom in the prior or how informative the prior is. When $d$ is large, the random realization of $V$ will be closer to $\tilde V$, i.e., the prior plays a more important role in the posterior. Since we want to infer the mean of $V$ more from the data, we use the smallest possible integer, one, to make the priors less informative. Vague priors for all other parameters are proposed as $\tau_H, \tau_F\sim Ga(10^{-3},10^{-3})$,  $\gamma_H, \gamma_F\sim Unif(0,10^6)$, $\beta\sim N(0,10^6)$, $\nu_H,\nu_F\sim Ga(10^{-3},10^{-3})$, $\phi_H, \phi_F\sim IG(10^{-3},10^{-3})$, $\gamma_{Hm}, \gamma_{Fm}\sim Unif(0,10^6)$, $\tau_W\sim Ga(10^{-3},10^{-3})$, where $IG$ stands for the inverse Gamma distribution and $Unif$ stands for the Uniform distribution.

To estimate the posterior distribution of all the latent states and parameters, we use Markov Chain Monte Carlo (MCMC) with Gibbs sampling and Metropolis-Hasting for parameters without an analytic form of the marginal posterior distribution. All the formulae for the Gibbs and Metropolis-Hasting updates are given in Section~\ref{sup:formulae} in the Supplementary Materials.
The constant $\kappa$ that reflects how inadequate the climate models represent the actual earth system is not identifiable in this Bayesian hierarchical setup. Due to our lack of knowledge about adequacy of the climate models, we assume $\kappa=1$ in the simulation and application studies in Sections~\ref{sec:simulation} and \ref{sec:application}, the value of which was also used in \cite{Sansom2017}. Domain experts may have insights into other choices of $\kappa$ values which may improve the inference of the future climate,  if the specified value is more reflecting the true inadequacy. However, we simply use $\kappa=1$ for the most general case. 
	
	\section{Simulation study}\label{sec:simulation}
In Section~\ref{sec:model}, we have proposed a new Bayesian hierarchical model, the novelty of which is that both the climate model dependence and the spatial correlation are well accounted for. 
In order to investigate the properties of this model and the validity of our inference procedures, we conduct extensive simulations by generating synthetic data for which the underlying model is known. We test different aspects of the proposed model and gain insights into what part of the model is worth modeling and what model components we may not be able to estimate even if they exist.
Careful simulation studies are particularly necessary for a complex model like the one we proposed because parameters in certain layers may not be estimated very well, and we need to investigate whether this would affect the estimation of variables of our primary interest.

\begin{figure}[ht!]
	\centering
	\includegraphics[width=\linewidth]{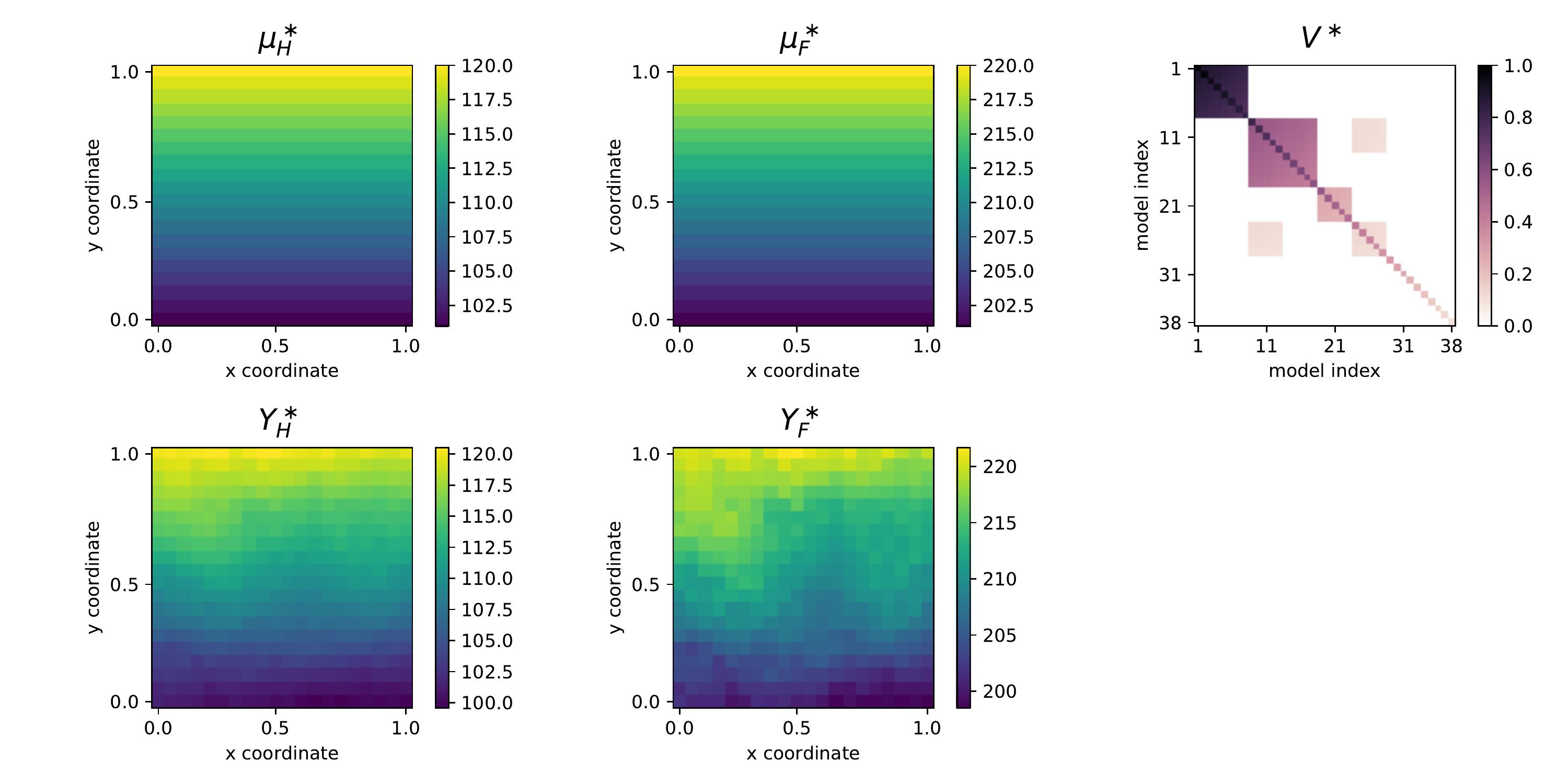}
	\caption{Top panels show the fixed consensus random fields $\mu_H^\ast$ and $\mu_F^\ast$ in the historical and future periods, respectively, and the model dependence matrix $V^\ast$ used to generate the synthetic data in the simulation study. Bottom panels give a realization of the randomly generated expected climate $Y_H^\ast$ and $Y_F^\ast$ in the historical and future periods, respectively.}
	\label{fig:Full-Design}
\end{figure}

\begin{figure}[ht!]
	\centering
	\includegraphics[width=\linewidth]{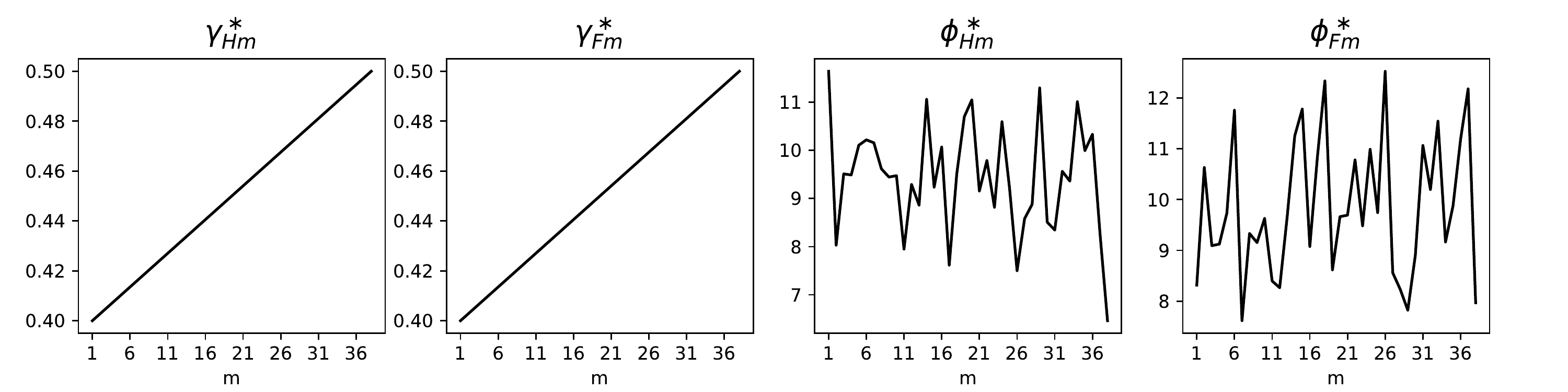}
	\caption{The fixed range parameters $\gamma_{Hm}^\ast$, $\gamma_{Fm}^\ast$ and one realization of the randomly generated inverse sill parameters $\phi_{Hm}^\ast$, $\phi_{Fm}^\ast$ for $m=1,\ldots,M$ in the simulation design.}
	\label{fig:Full-Design-Contd}
\end{figure}

In the simulation study, we run experiments with the following parameter setting.
We choose $n=20\times20=400$ locations over a regular grid in the two-dimensional domain $[0,1]\times[0,1]$.
The number of climate models is $M=38$, each of which has $R_{Hm}=R_{Fm}=10$ model runs. The number of observations data sets is $N=5$.
The true values of the consensus fields $\mu^\ast_H$, $\mu^\ast_F$ and the climate model dependence matrix $V^\ast$ used to generate the synthetic data are chosen as shown in Figure~\ref{fig:Full-Design}. The chosen values of $V^\ast$ make most pairs of climate models independent, while several clusters of climate models with correlation ranging from strong to weak exist. The expected climate is a latent state that is randomly generated from Formula~(\ref{eq:expectedClimate}), and one realization as an example is also shown in Figure~\ref{fig:Full-Design}.
We fix the true parameter values as $\gamma^\ast_H=0.5,\gamma^\ast_F=0.5,\tau^\ast_H=1.5,\tau^\ast_F=2,\tau^\ast_W=2,\beta^\ast=2,\phi^\ast_H=10,\phi^\ast_F=10,\nu_H^\ast=100,\nu_F^\ast=100,\phi^\ast_{Ha}=10,\phi^\ast_{Fa}=10$.
The selected true values $\gamma_{Hm}^\ast,\gamma_{Fm}^\ast$ for $m=1,\ldots,M$ are shown in Figure~\ref{fig:Full-Design-Contd} while the true values $\phi_{Hm}^\ast,\phi_{Fm}^\ast$ for $m=1,\ldots,M$ are randomly generated from the Gamma distributions specified in Formula~\eqref{eq:phiHmFm} based on the chosen $\nu_H^\ast,\nu_F^\ast,\phi_H^\ast,\phi_F^\ast$, one realization of which is also shown in Figure~\ref{fig:Full-Design-Contd}. 
Note that the choice for $\gamma_{Hm}^\ast,\gamma_{Fm}^\ast$ is arbitrary, which allows us  to cover a range of values corresponding to moderate spatial correlation that vary for different models. 

\begin{figure}[ht!]
	\centering
	\includegraphics[width=\linewidth]{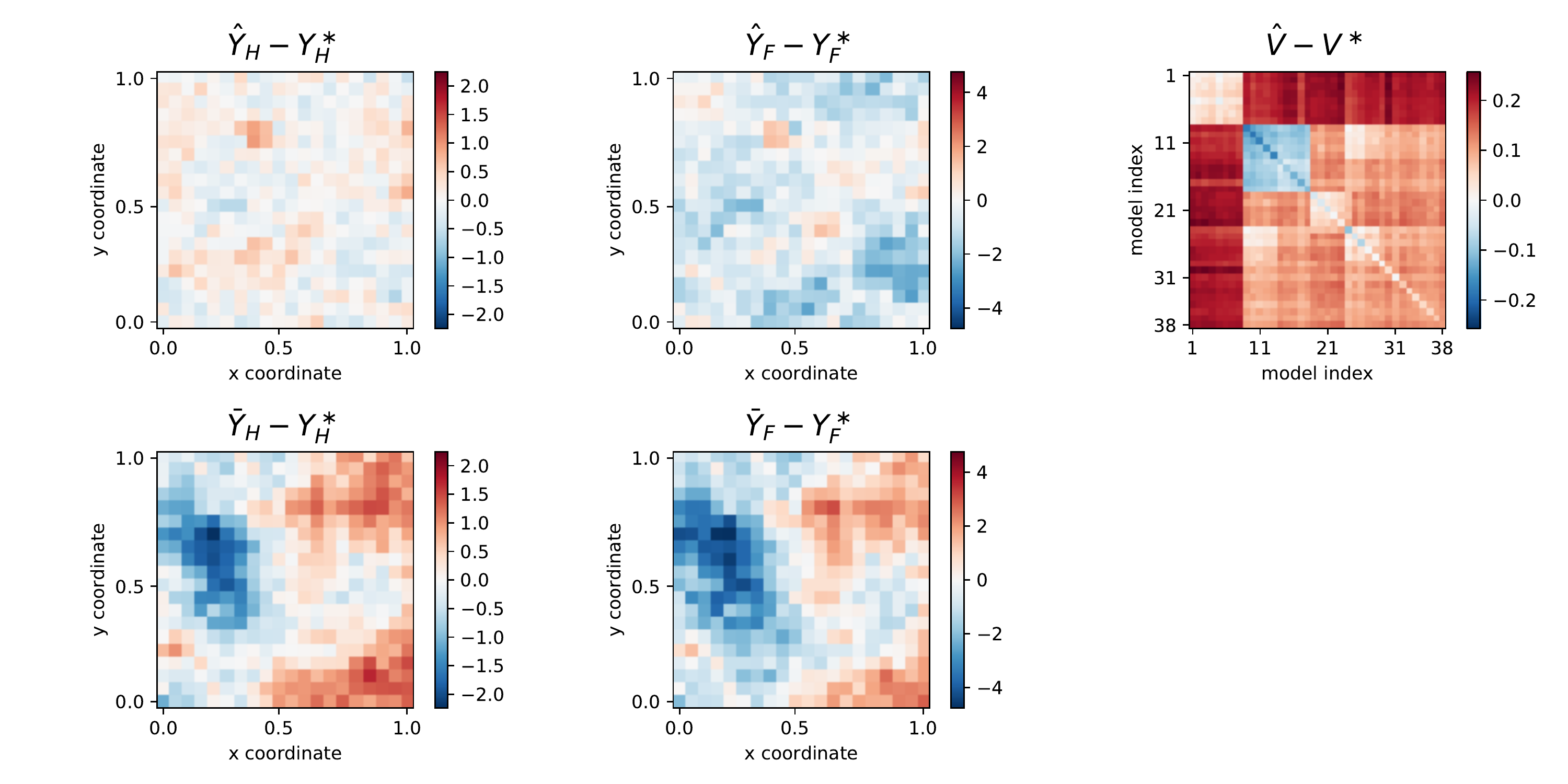}
	\caption{Differences between the estimates and the true values of $Y_H$, $Y_F$, and $V$. The marker $\hat\ $ represents the posterior means in the MCMC using our proposed Bayesian hierarchical model; the marker $\bar\ $ represents the multi-model means calculated as the averages of all the climate model runs; the superscript $^\ast$ represents the true values.}
	\label{fig:Full-Field}
\end{figure}

\begin{figure}[t!]
	\centering
	\includegraphics[width=\linewidth]{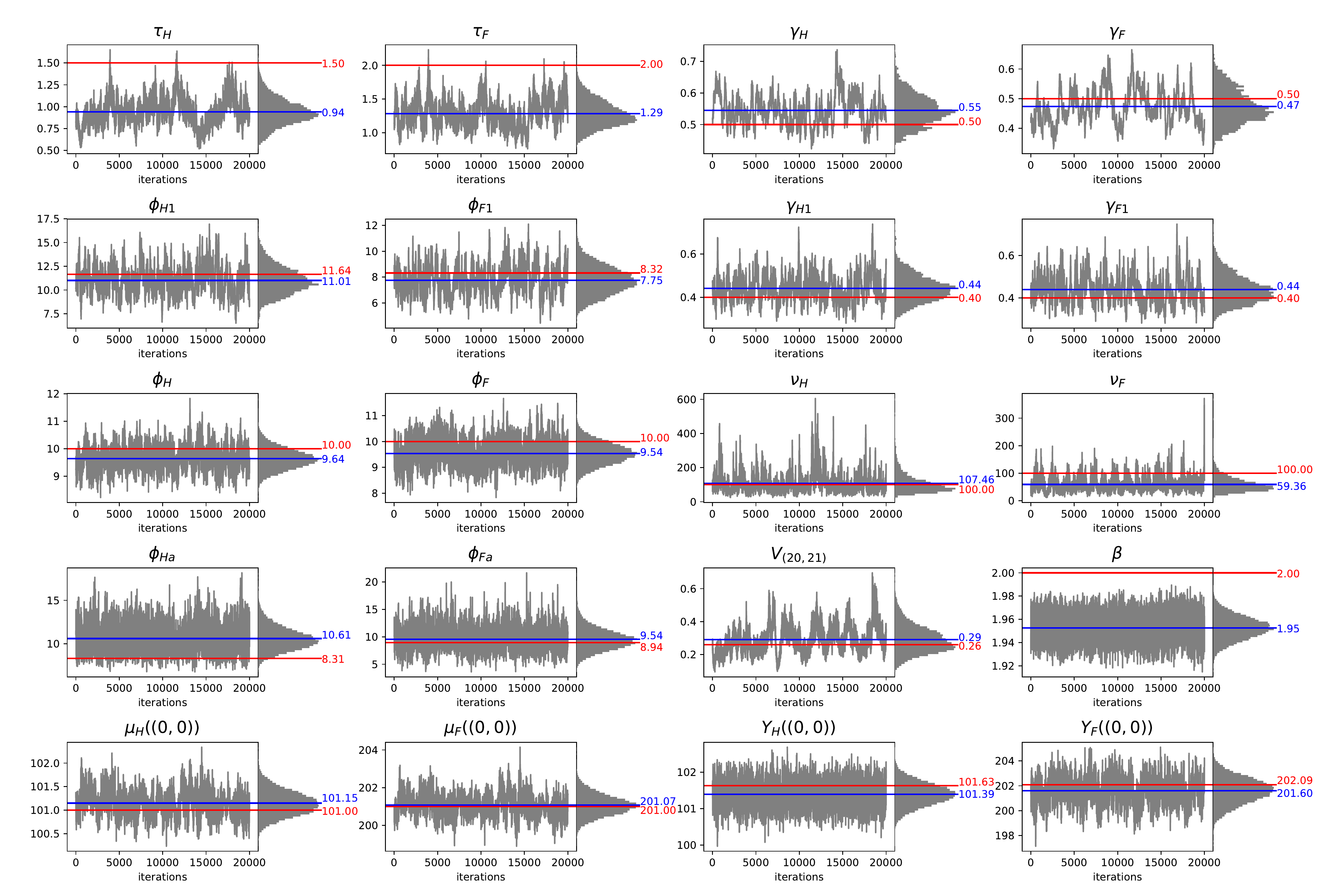}
	\caption{The trace plots, the histograms, the posterior means (blue), and the true values (red) of different parameters or latent states in the MCMC. Note that in the bottom panels, we only show the trace plots of some latent states at location $\bs=(0,0)$.}
	\label{fig:Full-Results}
\end{figure}

We estimate all the parameters or latent states in the Bayesian hierarchical model through an MCMC with 30,000 iterations, where the first 10,000 iterations are considered burn-in and discarded. The prior mean of the climate model dependence matrix, $\tilde V$, is chosen as an identity matrix because we try to make the prior as non-informative as possible and thus assume the prior has no knowledge about the correlation at all.
The posterior mean of each parameter or latent state in the last 20,000 iterations is used as the posterior estimate. We examined multiple independent MCMC runs with different appropriate initial values, and the difference among their results are subtle. We show in Figure~\ref{fig:Full-Field}
the differences between the posterior estimates and the true values of variables of our primary interest---the historical and future expected climate $Y_H$ and $Y_F$, as well as the climate model dependence matrix $V$, from one randomly selected MCMC run.
Since the multi-model mean, which is the straight average of all available climate model outputs, is commonly used in forecasting the future climate in the IPCC report, we also provide the differences between the true expected climate and the multi-model mean in Figure~\ref{fig:Full-Field}.
It can be observed that our posterior estimates lead to more accurate values in both discovering the historical expected climate and forecasting the future expected climate. Furthermore, since our inference is from a Bayesian model, it is very convenient to characterize the uncertainty of the expected climate, which, however, is challenging for the commonly used multi-model mean estimates. Uncertainty quantification is particularly important for making probabilistic forecasts. The estimated posterior distributions through the 20,000 MCMC iterations are shown in Figure~\ref{fig:Full-Results}. For most parameters or latent states, we observe that the estimated posterior distributions are Gaussian-like and the posterior means are accurate estimates, although the estimation of certain parameters such as $\tau_H$, $\tau_F$, $\nu_H$, and $\nu_F$ is less satisfactory. The latter is not surprising as it is common that some parameters are difficult to estimate in a complex Bayesian hierarchical model. More importantly, the expected climate $Y_H$ and $Y_F$, which are of our primary interest, have superior posterior estimates, regardless of a few other poorly estimated parameters.

\begin{figure}[t!]
	\centering
	\includegraphics[width=\linewidth]{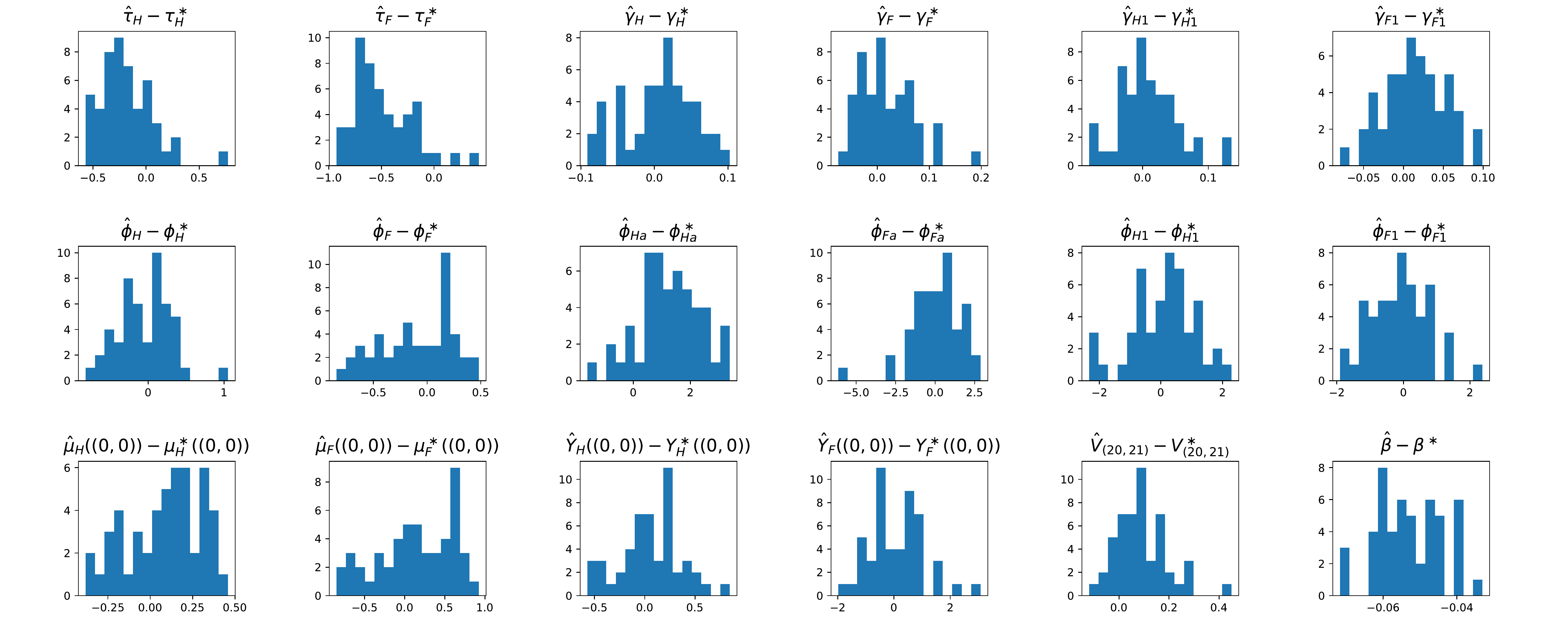}
	\caption{The histograms of the differences between the posterior estimates of parameters or latent states and the true values in the 50 independent experiments. The marker $\hat\ $ represents the posterior mean using our proposed Bayesian hierarchical model in the MCMC; the superscript $^\ast$ represents the true values. Note that in the bottom panels, we only show the histograms of some latent states at location $\bs=(0,0)$.}
	\label{fig:Full-Multiple-Runs}
\end{figure}

To have more assessment of the estimation performance, we show MCMC results from 50 generated synthetic data sets based on the same parameter setting ($\phi^\ast_{Hm}$ and $\phi^\ast_{Fm}$ are randomly generated for each data set and may be different). The histograms of the 50 differences between the posterior estimates and the true values for several parameters and latent states are shown in Figure~\ref{fig:Full-Multiple-Runs}.
We notice that $\beta$ tends to be slightly underestimated. We believe the underestimation is partially caused by the relatively small number of climate model runs. To verify our conjecture, we conducted experiments where a reduced number of climate model runs and observation data sets are used, in accordance to the actual number of climate model runs in CMIP5 and the observation (reanalysis) data sets discussed in Section~\ref{sec:application}. We found that the underestimation of $\beta$ is more significant in such case; however, the estimation of the expected climate is still accurate. Details of this investigation are provided in Section~\ref{sup:reality} in the Supplementary Materials. The overall conclusion is that the estimation of $\beta$ may be less reliable if the number of climate model runs is small, but the estimate for the expected climate remains intact. If learning the emergent relationship $\beta$ is of special interest, a large number of climate model runs are required.

\begin{figure}[ht!]
	\centering
	\includegraphics[width=\linewidth]{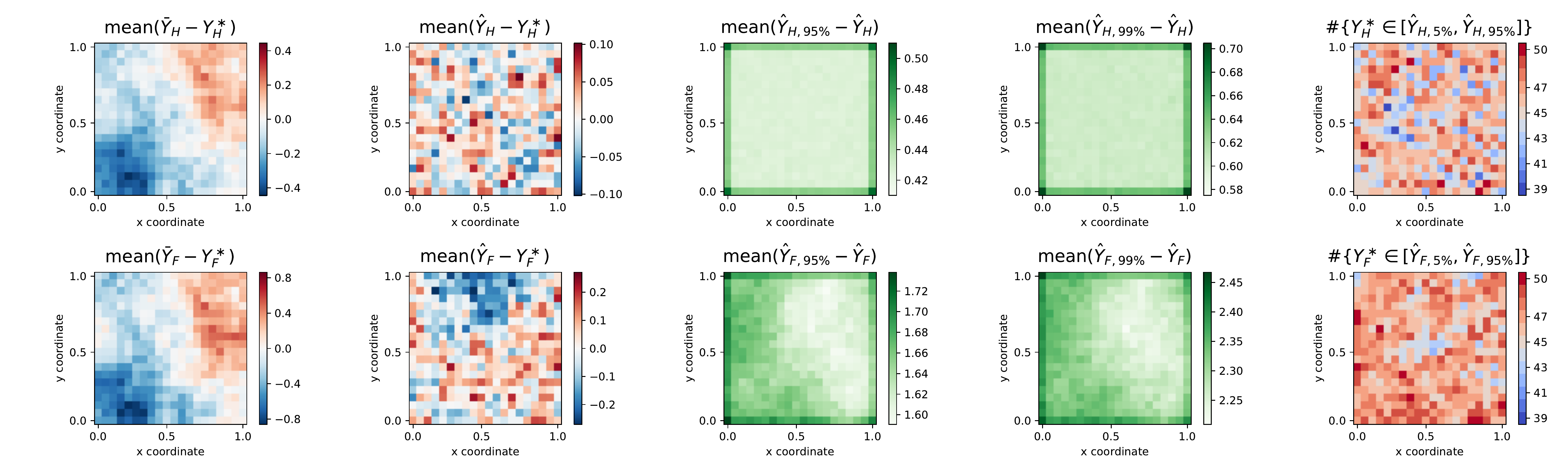}
	\caption{The first column shows the sample mean of the differences between the multi-model means ($\bar Y_H$, $\bar Y_F$) and the true values ($Y_H^\ast$, $Y_F^\ast$) in the 50 independent experiments. The second columns shows the sample mean of the differences between the posterior means ($\hat Y_H$, $\hat Y_F$)  and the true values. The third and forth columns show the sample mean of the differences between the posterior $95\%$-quantiles ($\hat Y_{H,95\%}$, $\hat Y_{F,95\%}$) or the posterior $99\%$-quantiles ($\hat Y_{H,99\%}$, $\hat Y_{F,99\%}$) and the posterior means. The fifth column shows the number of cases out of the 50 experiments whose true values fall into the $90\%$ credible intervals.}
	\label{fig:Full-Field-Multiple-Runs}
\end{figure}

The sample mean of the differences between the posterior estimates and true values of the whole expected climate fields $Y_H$ and $Y_F$ in the 50 experiments are shown in Figure~\ref{fig:Full-Field-Multiple-Runs}.
For comparison, we also show in Figure~\ref{fig:Full-Field-Multiple-Runs} the sample mean of the differences between the multi-model mean estimates and the true values in the 50 independent experiments, where we see our results yield much more accurate estimates.
Our estimates of unknown parameters and latent states in the 50 experiments in general perform very well, especially for the future expected climate $Y_F$, which is the quantity we are mostly interested in and trying to forecast.
Examining the accuracy of credible intervals allows us to assess the performance of the variability estimation; we therefore provide the $95\%$- and $99\%$-quantiles in the posterior distribution and the number of experiments whose true values fall into the estimated $90\%$ credible intervals (the interval between the $5\%$- and $95\%$-quantiles) in Figure~\ref{fig:Full-Field-Multiple-Runs}. The $95\%$- and $99\%$-quantiles are treated as the moderate and severe extremes, respectively.
An artifact is observed that the variability along the area border is generally larger than in the interior area because for the random process at locations along the border, there are fewer other moderately-correlated locations. If the variability of some locations along the border is of interest, this artifact can be easily resolved by expanding the area slightly. Comparing the number of experiments whose true values fall into the $90\%$ credible intervals to the theoretical benchmark, 45, which is $90\%$ of the 50 experiments, we conclude that the variability of the estimated expected climate is precisely quantified.

To demonstrate the importance of including spatial correlation and climate dependence in the Bayesian hierarchical model, we  conducted experiments where these components are ignored. The model that ignores the spatial correlation and climate model dependence acts like the model proposed by \cite{Sansom2017} for combining multiple climate model runs. Using this simplified model, we found a larger bias in the posterior mean of the expected climate as well as less accurate estimated variability, leading to a poorly estimated credible interval. 
This demonstrates the necessity of taking these two types of correlations into account in the Bayesian hierarchical model, observing the important role these correlations played in making inference of the expected climate. 
Detailed results of all these investigations are given in Section~\ref{sup:simplifiedModels} in the Supplementary Materials.
Noting that the study areas in the application in Section~\ref{sec:application} have a smaller number of locations than what we have in the simulation study, we also conducted experiments on the synthetic data with the same number of locations as in Section~\ref{sec:application}. We found that the results are quite similar and for succinctness, we do not present these additional experiments.

\section{Application}\label{sec:application}
Near-surface air temperature plays an important role in climate research and is a common output in many climate model products. We use the near-surface air temperature fields from 38 climate models with a total number of 81 climate model runs from CMIP5. This climate model data was also used by \citet{Herger2018}. The number of available model runs for each model is given in Table~\ref{tab:CMIP5} in the Supplementary Materials; the table also shows the model indices that we arbitrarily assigned.

When climate models construct projections for future climate change, it is necessary to make assumptions about future emission patterns and their consequences for greenhouse gas levels in the atmosphere. The IPCC has treated this issue by formulating several Representative Concentration Pathways (RCPs) that are trajectories of greenhouse gas concentrations adopted for its Fifth Assessment Report in 2014.  The major pathways used for climate modeling are RCP2.6, RCP4.5, RCP6, and RCP8.5, where the labels refer to possible ranges of radiative forcing values in the year 2100 (in watts per square meter). In this paper, we use RCP4.5 and RCP8.5. Loosely, RCP4.5 corresponds to a moderate degree of emission control in which greenhouse gases peak around 2040 and then decline. On the other hand, RCP8.5, often referred to colloquially as the ``business as usual" scenario, assumes that emissions will continue to increase throughout the twenty-first century.

We have historical values simulated by climate models from the year 1956 to 2013 and forecasted future values from 2006 to 2100 under RCP4.5 or RCP8.5 forcings. We truncate the periods to an equal length for the past and future as 1971--2000 and 2071--2100, respectively; we choose these periods to eliminate the overlap between the historical and future periods, to avoid the potentially less trustworthy simulation results in the first several years in the climate model products, and to put emphasis on the long-term forecast. 

For the observation data sets, we use two reanalysis data sets, which were also used in \citet{Herger2018}: Berkeley BEST Land (\url{http://berkeleyearth.org/data/}) and CRU TS (\url{https://crudata.uea.ac.uk/cru/data/hrg/cru_ts_3.23/}). There are three more reanalysis data sets studied in \citet{Herger2018}. However, two of them have lower resolution, and another one is the Berkeley BEST Global data, which is highly correlated with the Berkeley BEST Land data. Hence, we exclude those three data sets in our analysis.

\begin{figure}[t!]
	\centering
	\includegraphics[width=\linewidth]{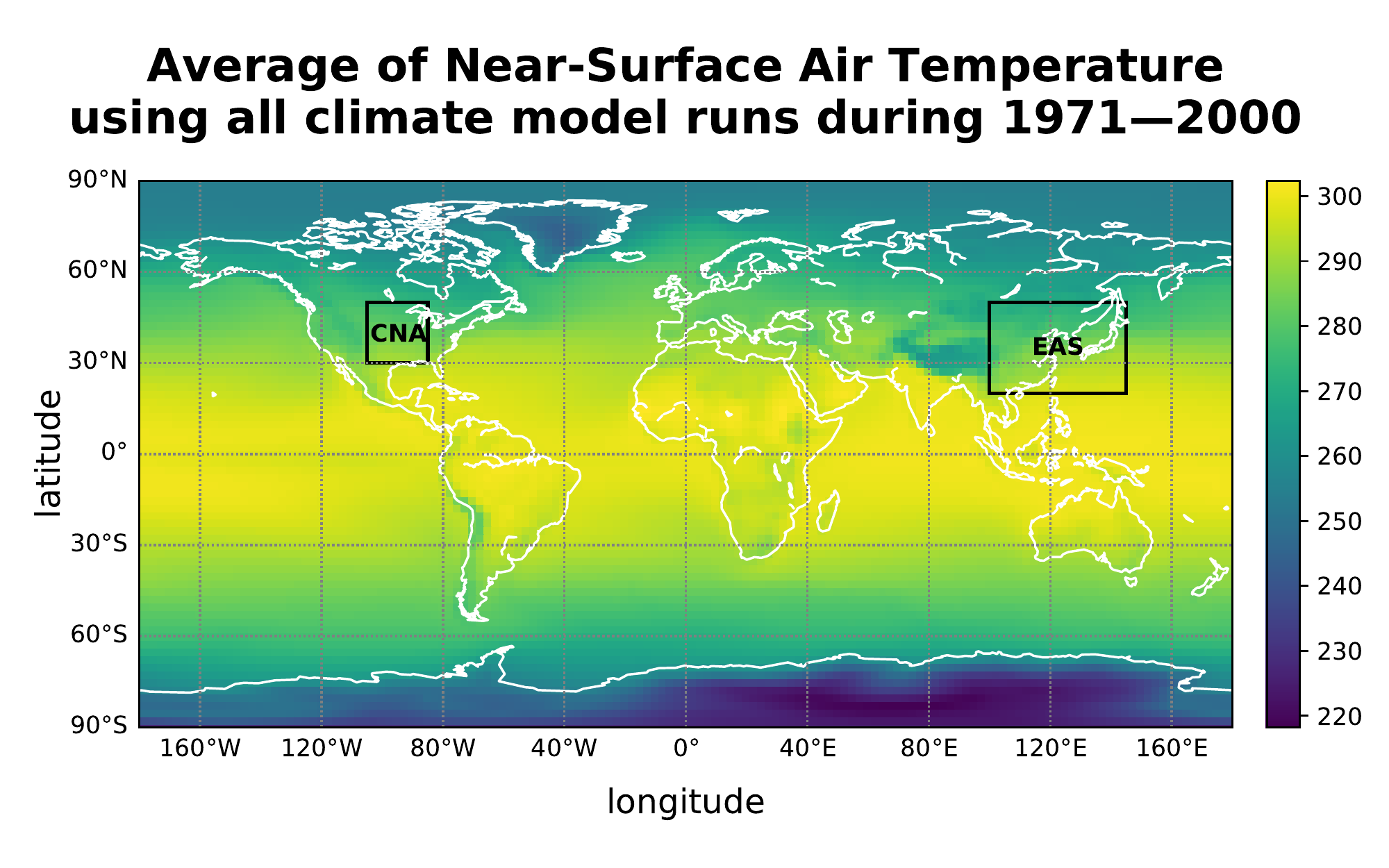}
	\caption{The average of the near-surface temperature (in kelvins) in all climate model runs during 1971--2000 over the globe.}
	\label{fig:ShowMap}
\end{figure}

\begin{figure}[t!]
	\centering
	\includegraphics[width=\linewidth]{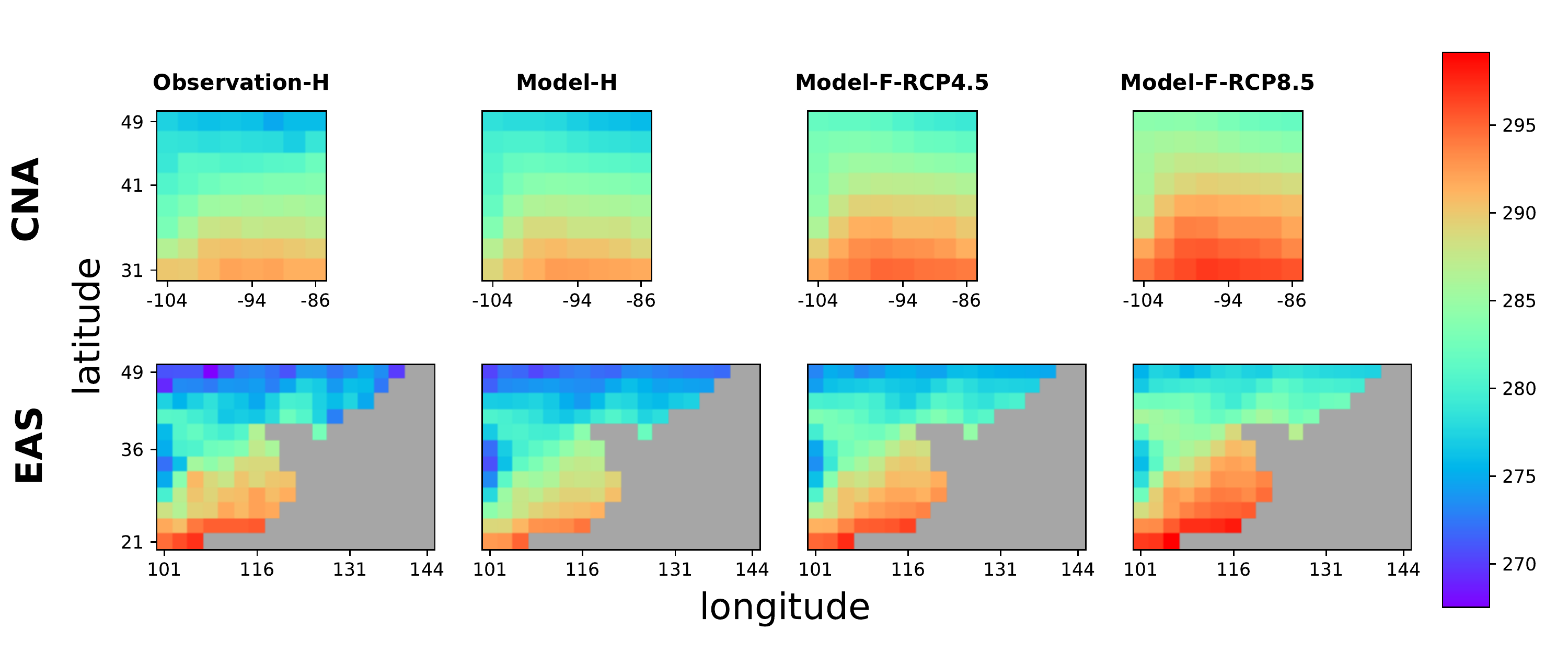}
	\caption{The average of the near-surface temperature (in kelvins) in the two reanalysis data sets during 1971--2000 (``Observation-H''), in all the climate model runs during 1971--2000 (``Model-H'') and during 2071--2100 under RCP4.5 (``Model-F-RCP4.5'') or RCP8.5 (``Model-F-RCP8.5'') in the Central North America (CNA) region and the East Asia region excluding islands (EAS).}
	\label{fig:regions}
\end{figure}

We investigate the near-surface air temperature in two regions with different characteristics: the Central North America (CNA) region and the East Asia (EAS) region, illustrated in Figures~\ref{fig:ShowMap} and \ref{fig:regions}. Figure~\ref{fig:ShowMap} shows the average of the near-surface temperature during 1971--2000 over the entire globe using all climate model outputs; Figure~\ref{fig:regions} shows the average of the observation data sets and climate model runs during 1971--2000 as well as 2071--2100 under RCP4.5 or RCP8.5 in the two study regions. For the EAS region, because a single spatial covariance model such as the stationary isotropic Whittle covariance function may not adequately represent the dependence between the continent and island locations, we focus specifically on mainland temperatures and thus exclude Japan for instance. The four application cases to investigate are the near-surface air temperature in  CNA  under RCP4.5, in EAS under RCP4.5, in CNA under RCP8.5, and in EAS under RCP8.5. In each case, we run MCMC with 130,000 iterations. The first 30,000 iterations are used as burn-in and discarded. We choose every $5^\text{th}$ value in the remaining 100,000 iterations to reduce auto-correlations existing in consecutive iterations in the MCMC. After thinning, we have 20,000 samples representing the posterior distribution.

\begin{figure}[t!]
	\centering
	\includegraphics[width=\linewidth]{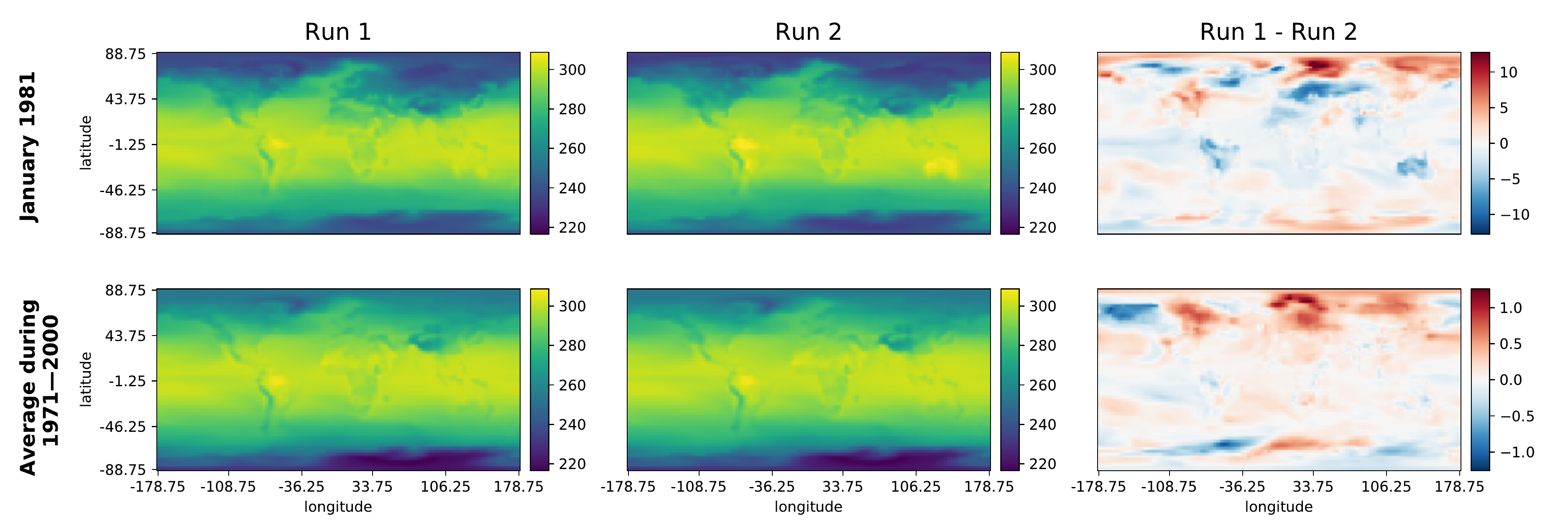}
	\caption{Two model runs and their differences (in kelvins) from the climate model CanESM2. Top panels show the monthly average of the near-surface air temperature in January 1981, and bottom panels show the average during the historical period from 1971 to 2000.}
	\label{fig:applicationData}
\end{figure}

In each application case, we compare our results with the multi-model mean, which is the average of all the available climate model runs used in the current IPCC report. In addition, \cite{Sansom2017}, hereafter SSB, proposed a model discussed in Section~\ref{sec:intro} and studied as a simpler version of our model (without accounting for spatial correlation and climate model dependence) in the simulation study with results provided in Section~\ref{sup:simp} in the Supplementary Materials. More specifically, in the SSB model, Formulas~\eqref{eq:covForClimateModels} and \eqref{eq:modelRuns} are modified by removing the spatial correlation $c(\|\bs_i-\bs_j\|)$ and the climate model dependence $v_{pq}$ as follows,
\[
\begin{array}{ccccc}
\cov\big(\epsilon_{Hp}(\bs_i),\epsilon_{Hq}(\bs_j)\big)&:&{\tau\inv_H}c(\|\bs_i-\bs_j\|;\gamma_H)v_{pq}&\longrightarrow&\tau_H\inv,\\
\cov\big(\epsilon_{Fp}(\bs_i),\epsilon_{Fq}(\bs_j)\big)&:&{\tau\inv_F}c(\|\bs_i-\bs_j\|;\gamma_F)v_{pq}&\longrightarrow&\tau_F\inv,\\
\cov\big(\epsilon_{Hmr}(\bs_i),\epsilon_{Hmr}(\bs_j)\big)&:&\phi_{Hm}\inv c(\|\bs_i-\bs_j\|;\gamma_{Hm})&\longrightarrow&\phi_{Hm}\inv,\\ 
\cov\big(\epsilon_{Fmr}(\bs_i),\epsilon_{Fmr}(\bs_j)\big)&:&\phi_{Fm}\inv c(\|\bs_i-\bs_j\|;\gamma_{Fm})&\longrightarrow&\phi_{Fm}\inv.\\ 
\end{array}
\]
We clearly see the spatial correlation in the climate model runs in Figure~\ref{fig:applicationData}, where two model runs of CanESM2 in a specific month and the average over the historical period are shown. Looking at the differences depicted in the right panels in Figure~\ref{fig:applicationData}, we observe the spatial correlation in the deviation of climate model runs from the climate model means, because we would see white noises in the absence of spatial correlation.

\subsection{Results for the two regions under the two forcings}\label{subsec:applicationResults}

\begin{figure}[t!]
	\centering
	\includegraphics[width=\linewidth]{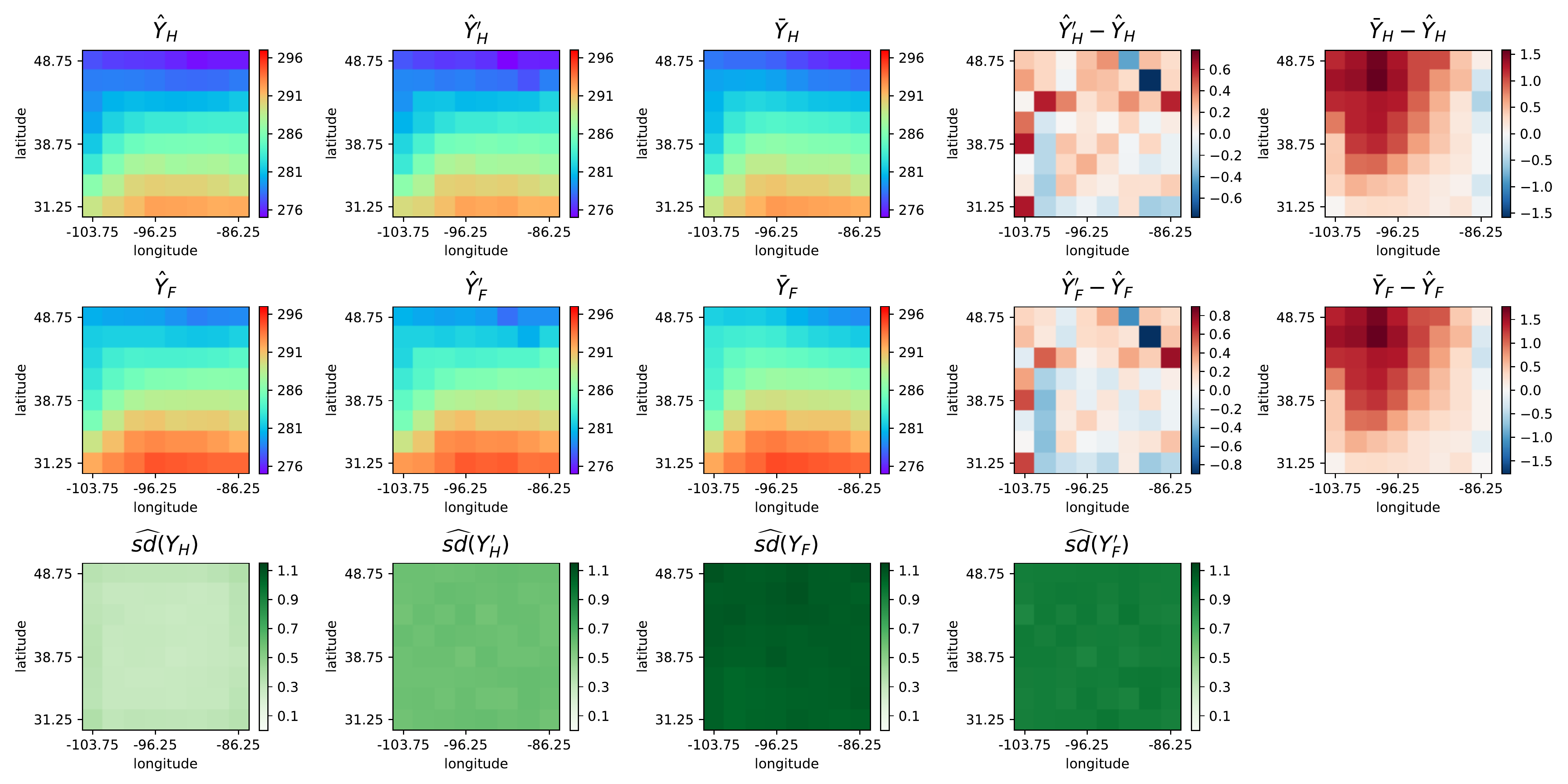}	
	\caption{Results in the  CNA region from different approaches, where the climate model outputs under RCP4.5 is used. $\hat Y_H$ and $\hat Y_F$ are the posterior mean of the expected climate in the historical and the future periods using our proposed hierarchical model, whereas $\hat Y_H^\prime$ and $\hat Y_F^\prime$ are the posterior mean using the SSB model. $\bar Y_H$ and $\bar Y_F$ are the multi-model mean, i.e., the average using all climate model runs. $\widehat{sd}(\cdot)$ is the estimated standard deviation in the MCMC. Unit: kelvin.}
	\label{fig:CNA-Field}
\end{figure}

Figure~\ref{fig:CNA-Field} summaries our posterior results compared to the SSB model and the multi-model mean using climate model outputs in the CNA region where RCP4.5 is used for the future forecast.
The multi-model mean provides much higher temperature estimates than our posterior mean at the majority of locations. The difference between the estimated temperatures from the SSB model and our model is comparatively small and has both positive and negative values. However, the SSB model yields a larger variability in the historical period and a smaller variability in the future than our model does.
Recall that in the simulation study (results given in Section~\ref{sup:simp} in the Supplementary Materials), we have shown that the SSB model fails to provide accurate mean and variability estimates.
Trace plots of selected parameters and latent states from two MCMC runs of our proposed Bayesian hierarchical model are given in Figure~\ref{fig:CNA-Trace} in the Supplementary Materials. Although the estimates of the spatial parameters, particularly those associated with the climate models with only one model run, are not optimal, the estimates of the most important latent states $Y_H$ and $Y_F$ are robust and trustworthy. This agrees with the findings in our simulation study where the number of climate model runs and observational products is identical to those in CMIP5 (detailed results are given in Section~\ref{sup:reality} in the Supplementary Materials).
The posterior distribution of the climate dependence matrix $V$ and the emergent relationship $\beta$ are also stable, but $\beta$ may be underestimated based on our findings in the simulation study. If learning the emergent relationship $\beta$ and all the spatial parameters is also of interest, more climate model runs are required, especially from the climate models that have only one model run (see Figure~\ref{fig:Reality-Spatial} in the Supplementary Materials). 

Figure~\ref{fig:RCP85-CNA-Field} shows the results in the CNA region where the climate model outputs under RCP8.5 are used for the future forecast.
\begin{figure}[t!]
	\centering
	\includegraphics[width=\linewidth]{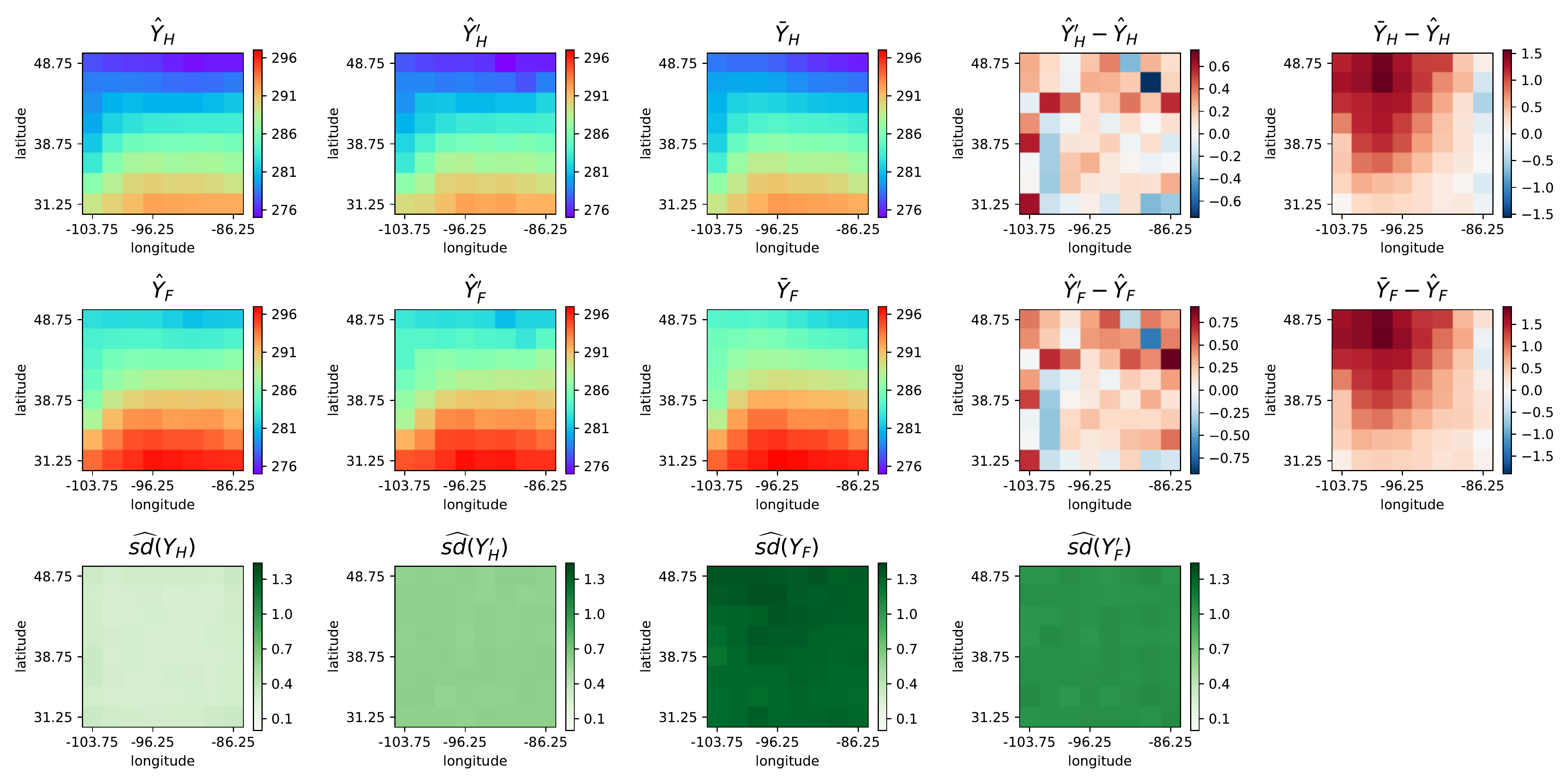}
	\caption{Results in the  CNA region from different approaches, where the climate model outputs under RCP8.5 are used. The same notation as in Figure~\ref{fig:CNA-Field} is used.}
	
	\label{fig:RCP85-CNA-Field}
\end{figure}
Compared to the results under RCP4.5, the future expected climate $Y_F$ for the near-surface temperature under RCP8.5 has higher values, which behaves as expected because of the higher emission level that contributes to global warming. The comparison between our model result and the multi-model mean or the SSB model is similar to that in the RCP4.5 case. The multi-model mean tends to provide much higher temperatures. The SSB model gives both higher and lower temperatures and yields a larger variability in the historical period but a smaller variability in the future.

\begin{figure}[t!]
	\centering

	\includegraphics[width=\linewidth]{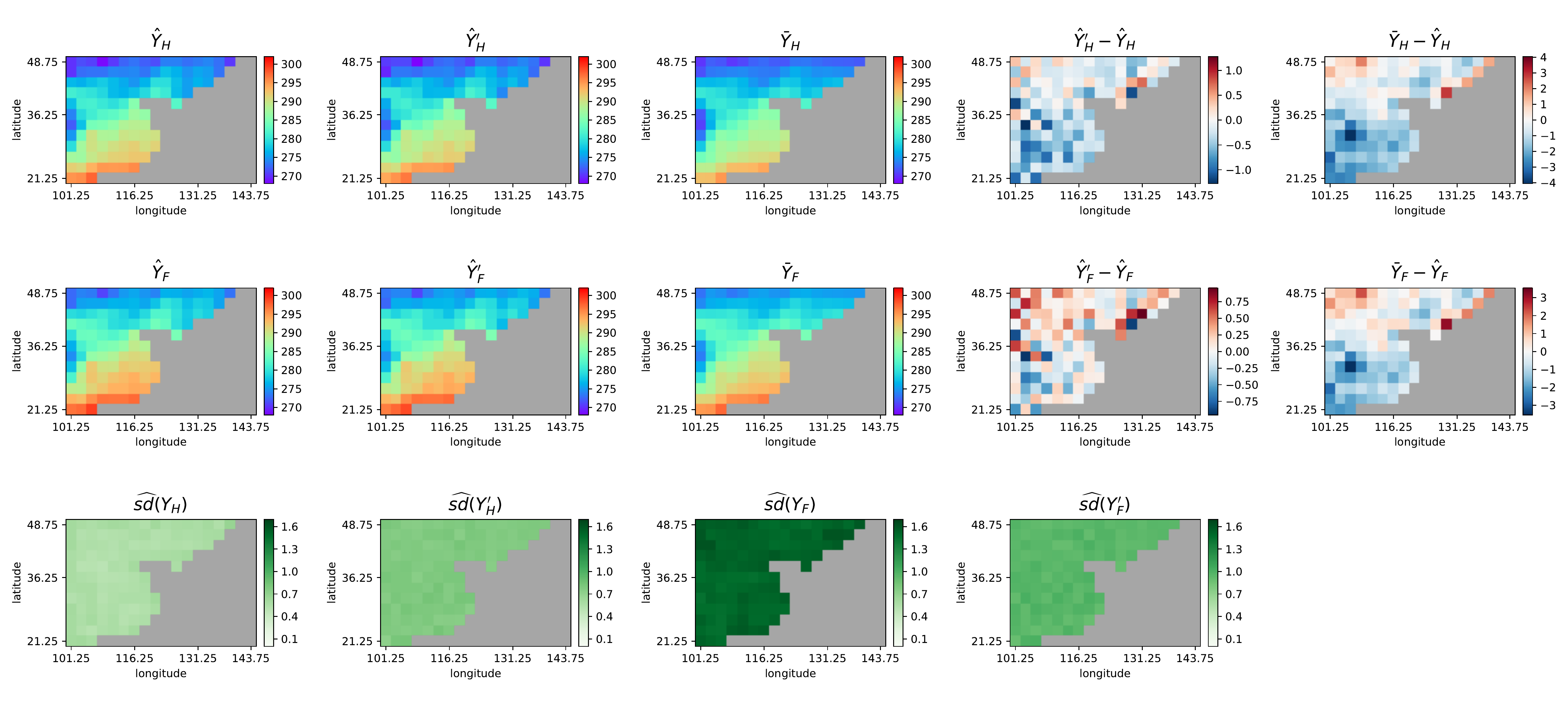}

	\caption{Results in the  EAS region from different approaches, where the climate model outputs under RCP4.5 are used. The same notation as in Figure~\ref{fig:CNA-Field} is used.}
	
	\label{fig:EAS-Field}
\end{figure}

\begin{figure}[ht!]
	\centering
	
	\includegraphics[width=\linewidth]{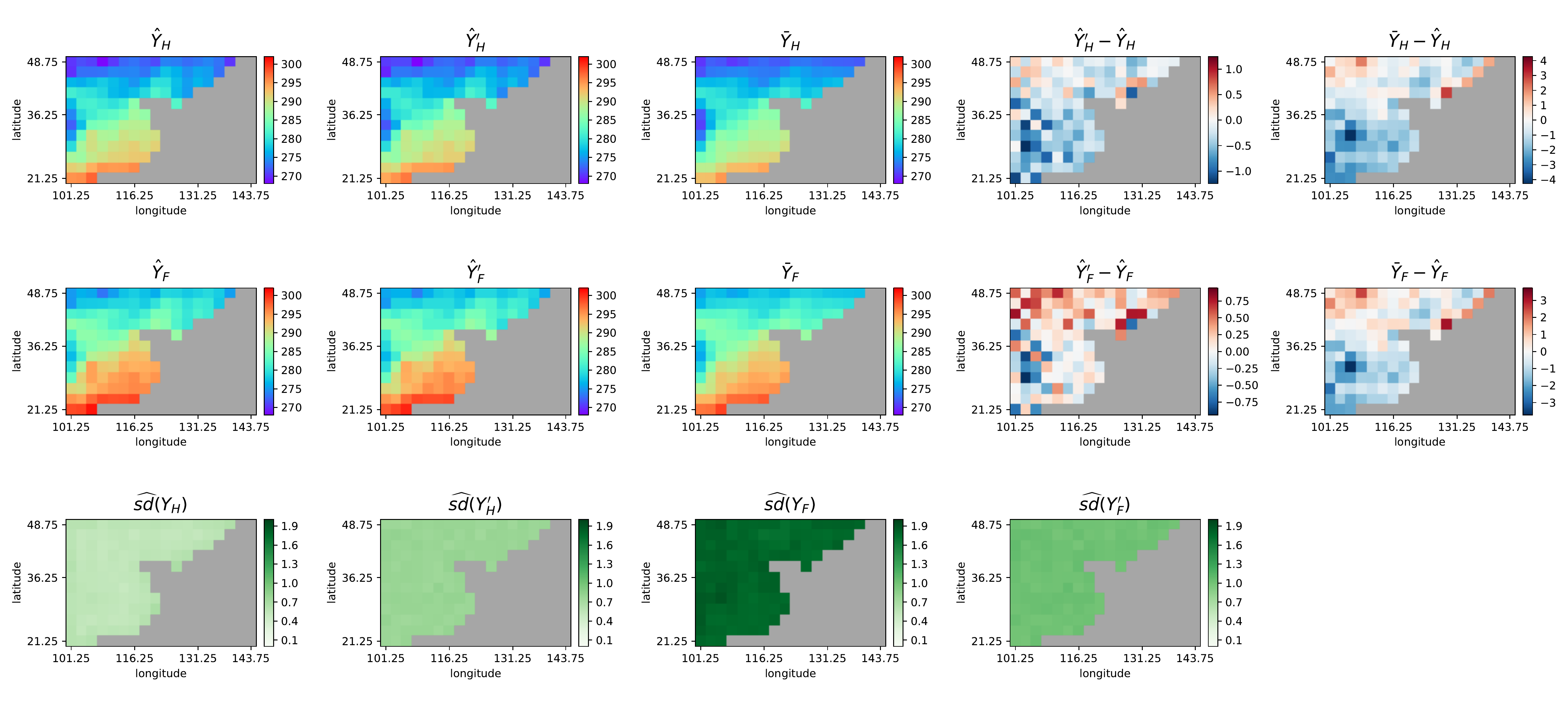}
	
	\caption{Results in the  EAS region from different approaches, where the climate model outputs under RCP8.5 are used. The same notation as in Figure~\ref{fig:CNA-Field} is used.}
	
	\label{fig:RCP85-EAS-Field}
\end{figure}

The inference results from different approaches in the EAS region are shown in Figure~\ref{fig:EAS-Field} under RCP4.5 and Figure~\ref{fig:RCP85-EAS-Field} under RCP8.5.
Most conclusions are similar to those in the CNA region except that the multi-model means now tend to have lower values at the majority of locations, especially in the southern part of the EAS region. We are not sure what causes the opposite signs of difference between the multi-model mean and our results in these two regions. One possible reason may be the different availability of observational records in these two regions that can be used to validate the climate models, but more scientific research is needed to interpret this finding.

\subsection{Probabilistic investigation of the difference between the multi-model mean and our approach}\label{subsec:twoForcings}
\begin{figure}[t!]
	\centering	
	\includegraphics[width=\linewidth]{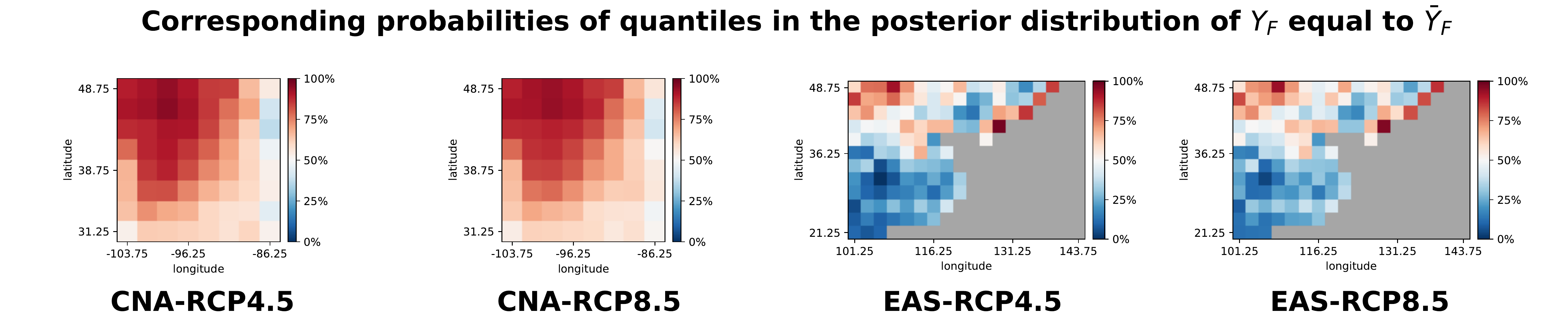}
	\caption{The corresponding probabilities of quantiles in the posterior distribution of the future expected climate $Y_F$ from our proposed model equal to the multi-model mean in each of the four application cases.}
	
	\label{fig:difference}
\end{figure}

In Section~\ref{subsec:applicationResults}, we purely discussed the difference between the multi-model mean and the posterior mean with our proposed model. However, the proposed Bayesian statistical model naturally yields a full posterior distribution, which allows us to examine the multi-model mean estimation in the context of the posterior distribution. Figure~\ref{fig:difference} shows the corresponding probabilities of the quantiles in the posterior distribution equal to the multi-model mean estimates for the future forecast, under both RCP4.5 and RCP8.5. We observe that at the majority of the locations in the CNA region under either forcing, the multi-model mean estimates correspond to very high quantiles in our posterior distribution, indicating large chances for overestimating the future near-surface temperature. However, in the EAS region, the multi-model mean estimates tend to underestimate the future near-surface temperature at most locations, especially in the southern part. The overestimation or underestimation patterns are quite similar under the two different forcings in the same region.

Table~\ref{tab:summary} further summarizes the climate impact of our work, where the differences of the predicted mean temperature for 2070--2100 over the entire CNA or EAS region between our method and the multi-model mean as well as the resulting $90\%$ credible intervals by our method are given. Our projected mean temperature in the CNA region for 2070--2100 is about 0.8 K lower than the multi-model mean, while in the EAS region it is about 0.5 K higher; however, in both cases, the widths of the $90\%$ credible intervals are of the order 3--6 K, so the uncertainties overwhelm the comparatively small differences in projected mean temperatures.
Similarly, we also provide these results for the SSB model inference. We see that the SSB model leads to higher mean temperatures than our model, and the widths of the $90\%$ are comparatively smaller, especially for the EAS region.

\begin{table}[ht!]
	\centering
	\caption{Differences of the predicted mean temperature for 2070--2100 over the entire CNA or EAS region between our method or the SSB model and the multi-model mean as well as the $90\%$ credible intervals resulting from our method or the SSB model. Unit: kelvin.}
	\begin{tabular}{c|c|c|c|c}
	\hline
	\multirow{2}{*}{Region and Forcing}& \multicolumn{2}{c|}{Region Mean Difference}&\multicolumn{2}{c}{$90\%$ Credible Interval of Region Mean}\\
	\cline{2-5}
	&~~$\hat Y_F - \bar Y_F$~~&$\hat Y_F^\prime - \bar Y_F$&~~~~~~~~~~$\hat Y_F$~~~~~~~~~~&$\hat Y_F^\prime$ \\
	\hline
	CNA, RCP4.5&-0.71&-0.67&[285.02,288.42]&[285.16,288.21]\\
	\hline
	CNA, RCP8.5&-0.84&-0.65&[286.55,290.77]&[287.25,290.68]\\
	\hline
	EAS, RCP4.5&0.49&0.51&[281.06,286.22]&[282.10,285.24]\\
	\hline
	EAS, RCP8.5&0.42&0.48&[282.60,288.75]&[284.21,287.59]\\
	\hline
	\end{tabular}
	\label{tab:summary}
\end{table}

\subsection{Summary of climate model dependence results}
\begin{figure}[h!]
	\centering	
	\includegraphics[width=\linewidth]{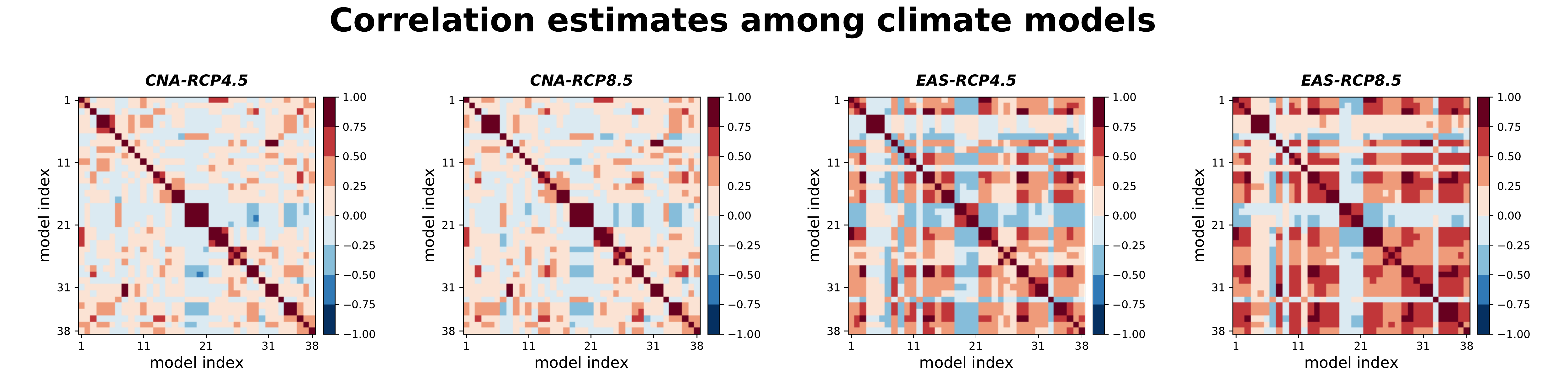}
	\caption{Correlation estimates among all the climate models for each of the four application cases.}
	\label{fig:V}
\end{figure}

A byproduct of this investigation is the posterior estimate of the climate model dependence matrix $V$. We normalize the posterior mean of $V$ to a correlation matrix and show the estimates for the four application cases in Figure~\ref{fig:V}. The estimated correlation matrices look similar under different forcings in the same region but look quite distinct in different regions. Generally speaking, the climate models are more correlated with each other in the EAS region.
One possible reason is that more observations in the CNA region can be used to calibrate climate models individually, leading to more independence among climate models in this region.

It is also observed that all the high correlations are positive, indicating the agreement of climate model outputs. Table~\ref{tab:V}
\begin{table}[ht!]
	
	\centering
	\caption{Climate model pairs with estimated correlations greater than $0.7$ in all four application cases. ``C45'' stands for the estimated correlation in the CNA region under RCP4.5, ``C85'' stands for the estimated correlation in the CNA region under RCP8.5, ``E45'' stands for the estimated correlation in the EAS region under RCP4.5, and ``E85'' stands for the estimated correlation in the EAS region under RCP8.5.}
	\begin{tabular}{cc|cccc}
		\hline
		Model 1 & Model 2  & C45 & C85 & E45 & E85\\
		\hline
		CCSM4 &  CESM1-BGC & 0.98 & 0.98 & 0.98 & 0.98 \\
		CMCC-CMS &  MPI-ESM-LR & 0.79 & 0.78 & 0.74 & 0.83 \\
		GFDL-ESM2G &  GFDL-ESM2M & 0.89 & 0.88 & 0.99 & 0.99 \\
		GISS-E2-H &  GISS-E2-H-CC & 0.93 & 0.93 & 0.97 & 0.97 \\
		GISS-E2-R &  GISS-E2-R-CC & 0.94 & 0.96 & 0.98 & 0.98 \\
		HadGEM2-AO &  HadGEM2-CC & 0.76 & 0.77 & 0.92 & 0.93 \\
		IPSL-CM5A-LR &  IPSL-CM5B-LR & 0.81 & 0.79 & 0.90 & 0.94 \\
		MIROC-ESM &  MIROC-ESM-CHEM & 0.88 & 0.94 & 0.98 & 0.99 \\
		MPI-ESM-LR &  MPI-ESM-MR & 0.94 & 0.94 & 0.93 & 0.95 \\
		NorESM1-M &  NorESM1-ME & 0.95 & 0.96 & 0.98 & 0.99 \\
		\hline
	\end{tabular}
	\label{tab:V}
\end{table}
lists the climate model pairs that have estimated correlations greater than $0.7$ in all four application cases. Except for the pairs \texttt{CCSM4} versus \texttt{CESM1-BGC} and \texttt{CMCC-CMS} versus \texttt{MPI-ESM-LR}, all the other highly correlated climate model pairs share the same main climate model but are coupled with different geophysical components. Both \texttt{CCSM} and \texttt{CESM} are climate models operated by National Center of Atmospheric Research in the USA, where the former is a subset of and has been superseded by the latter. Therefore, there is no surprise for this pair to be highly correlated, and as a matter of fact, this pair has the largest correlation among all the climate models. An interesting finding is the high correlation between \texttt{CMCC-CMS}, which is operated by the Euro-Mediterranean Center on Climate Change in Italy, and \texttt{MPI-ESM}, which is operated by the Max Planck Institute in Germany, although the correlation between this pair is comparatively lower than the other pairs. This high correlation may be due to the fact that both climate models make use of the atmospheric model component ECHAM. More domain expertise may be required to fully interpret the correlation between this climate model pair.

	\section{Discussion}\label{sec:discussion}

In this paper, we have extended previous approaches to multi-model ensembles by incorporating two features of climate models that have been analyzed on their own in previous papers, but not in conjunction with the other sources of variability in climate model projections: spatial correlation and dependence among climate models. As a result, we are able to produce posterior distributions for spatial climate model projections that incorporate natural and internal variability, biases and correlations in climate model outputs, emergent relationships, and the agreement of historical climate model runs with observational data. The results of Sections~\ref{subsec:applicationResults} and \ref{subsec:twoForcings} illustrate some comparisons between our approach and the earlier SSB approach, as well as the uniform model averaging approach.

There are still a number of limitations of our statistical model. In particular, it assumes that the spatial fields have a stationary isotropic structure and that the joint distributions of spatial fields over several models have a separable covariance structure, as is evident from Formula~\ref{eq:covForClimateModels}. These assumptions may be reasonable when applied to relatively small regions, but we would not expect a stationary isotropic spatial covariance function to be applicable over the whole earth. In addition, it would be worthwhile to investigate adding a temporal component to the model in order to accommodate inter-annual variability.

It could also be of interest to explore in more detail the potential of this approach to model climate extremes. Section~\ref{subsec:twoForcings} has shown how we can use quantiles of the posterior distribution to compare one set of model projections with another, but it would require a separate investigation to determine how robust the present approach is for calculating extreme quantiles of future climate variables.

As climate science moves from CMIP5 to CMIP6, there is likely to be even more demand for advanced statistical approaches for multi-model ensembles.

The code and data used in analyzing the near-surface temperature in the Central North America region and the East Asia region can be found in \url{https://github.com/hhuang90/Combine-CMIP5}.
	
	\if1\blind
	{
	\section*{\centering Acknowledgment}
    We thank Gab Abramowitz and Nadja Herger for providing the gridded near-surface air temperature data in CMIP5 and the reanalysis data sets. We thank Michael Wehner for providing information in interpreting our findings of highly-correlated climate model pairs. 
	}
	\fi

	\bibliography{ref}
	\bibliographystyle{chicago}
	
\pagebreak
\begin{center}
	\textbf{\LARGE Supplementary Materials}
\end{center}
\setcounter{section}{0}
\setcounter{equation}{0}
\setcounter{figure}{0}
\setcounter{table}{0}
\setcounter{page}{1}
\renewcommand{\theequation}{S\arabic{equation}}
\renewcommand{\thefigure}{S\arabic{figure}}
\renewcommand{\thetable}{S\arabic{table}}
\renewcommand{\thesection}{S\arabic{section}}
\renewcommand{\bibnumfmt}[1]{[S#1]}
\renewcommand{\citenumfont}[1]{S#1}

\section{Additional results in the simulation studies}\label{sup:sim}
In this section, we present in detail the additional results in the simulation studies when we ignore certain model parts in the inference procedure or reduce the number of climate model runs and observational data sets.

\subsection{Simplified model inference}\label{sup:simplifiedModels}
We show results of different scenarios of ignoring certain parts in the full Bayesian hierarchical model in the inference procedure.

\subsubsection{Absence of climate model dependence}\label{sup:NoV}
\begin{figure}[h!]
	\centering
	\includegraphics[width=\linewidth]{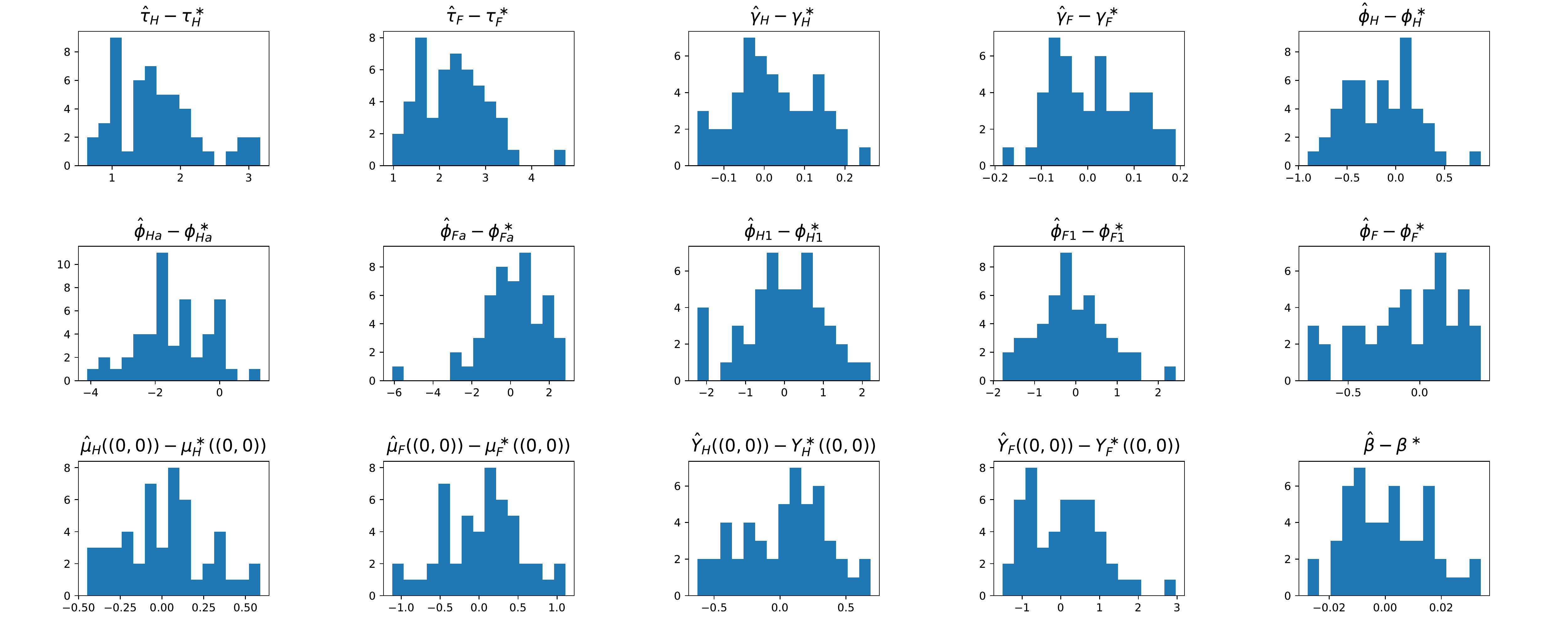}
	\caption{The histograms of the differences between the posterior estimates of parameters or latent states and the true values in the 50 independent experiments where we use Formula~\eqref{eq:NoV} for the random process covariance in the climate model means.  The marker $\hat\ $ represents the posterior mean using our proposed hierarchical model in the MCMC; the superscript $^\ast$ represents the true values.}
	\label{fig:NoV-Multiple-Runs}
\end{figure}

\begin{figure}[h!]
	\centering
	\includegraphics[width=\linewidth]{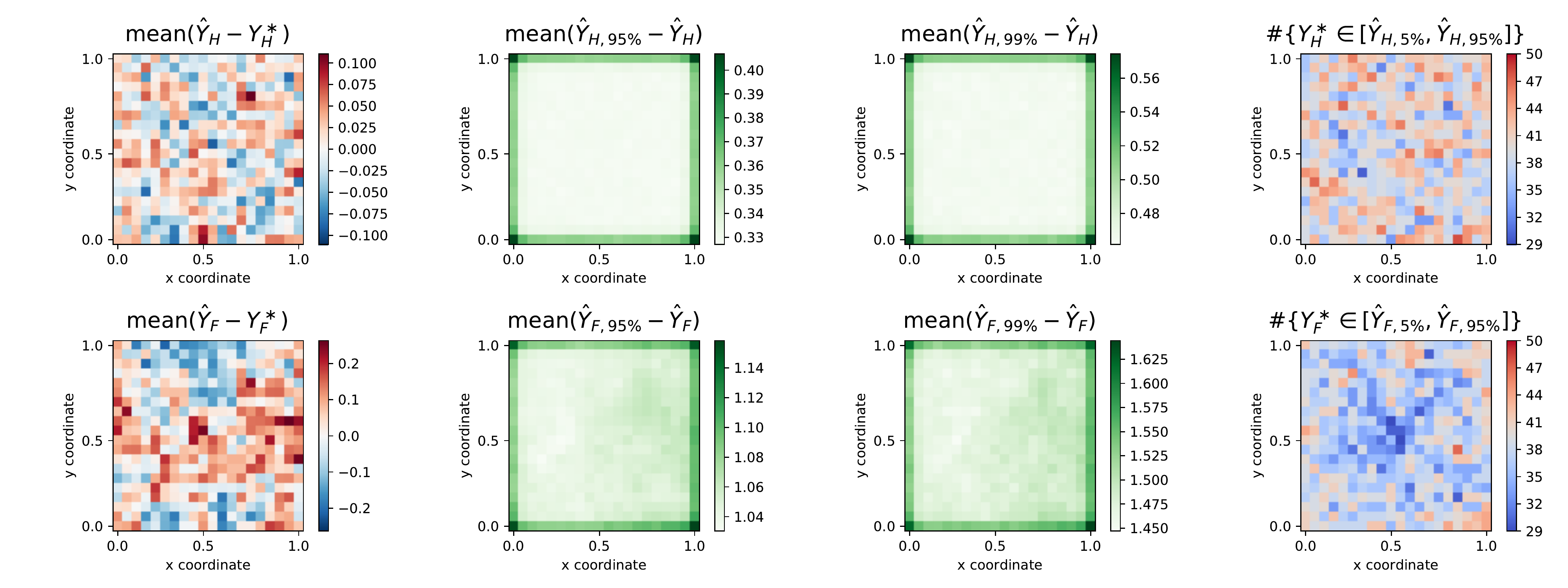}
	\caption{Summary of the results where we use Formula~\eqref{eq:NoV} for the random process covariance in the climate model means. The leftmost panels are the sample mean of the differences between the posterior means ($\hat Y_H$, $\hat Y_F$) and the true values ($Y_H^\ast$, $Y_F^\ast$) in the 50 independent experiments. The middle panels show the sample mean of the differences between the posterior $95\%$-quantiles ($\hat Y_{H,95\%}$, $\hat Y_{F,95\%}$) or the posterior $99\%$-quantiles ($\hat Y_{H,99\%}$, $\hat Y_{F,99\%}$)  and the posterior means in the 50 experiments. The rightmost panels show the number of cases out of the 50 experiments for the true values falling into the $90\%$ credible intervals.}
	\label{fig:NoV-Field-Multiple-Runs}
\end{figure}

The climate model dependence is ignored here, which means the covariance structure in Formula~\eqref{eq:covForClimateModels} is changed to Formula~\eqref{eq:NoV}.
\begin{equation}\label{eq:NoV}
\begin{array}{rcl}
\cov\big(\epsilon_{Hp}(\bs_i),\epsilon_{Hq}(\bs_j)\big)&=&{\tau\inv_H}c(\|\bs_i-\bs_j\|;\gamma_H),\\
\cov\big(\epsilon_{Fp}(\bs_i),\epsilon_{Fq}(\bs_j)\big)&=&{\tau\inv_F}c(\|\bs_i-\bs_j\|;\gamma_F).\\
\end{array}
\end{equation}
We also conducted 50 independent experiments using the same synthetic data as in Section~\ref{sec:simulation}, which is generated from the full model described in Section~\ref{sec:model} with true parameter values specified in Section~\ref{sec:simulation}. The posterior inference results are given in  Figures~\ref{fig:NoV-Multiple-Runs} and \ref{fig:NoV-Field-Multiple-Runs}. Compared to the full model inference results in Figure~\ref{fig:Full-Field-Multiple-Runs}, we see the errors of the posterior means are similar but the deviation of posterior $95\%$- and $99\%$-quantiles from the posterior means are much smaller, indicating an underestimated estimate variability. This is also demonstrated by the number of cases for the true values of $Y_H$ and $Y_F$ falling into the $90\%$ credible intervals where we may consider 45 as a theoretical benchmark.

\subsubsection{Absence of spatial correlation in climate model means in the inference}\label{sup:NoSp}
In this section, we ignore the spatial correlation in the climate model runs  by removing the spatially-correlated noises in Formula~\eqref{eq:modelRuns}  and using white noises shown in Formula~\eqref{eq:NoR} instead.
\begin{equation}\label{eq:NoR}
\begin{array}{rcl}
X_{Hmr}(\bs)&\sim& N\big(X_{Hm}(\bs),\phi_{Hm}\inv\big),\\
X_{Fmr}(\bs)&\sim& N\big(X_{Fm}(\bs),\phi_{Fm}\inv\big).
\end{array}
\end{equation}
In addition, the spatial correlation in the climate model means specified in Formula~\eqref{eq:covForClimateModels} is also ignored where we change the covariance structure to Formula~\eqref{eq:NoSp}.
\begin{equation}\label{eq:NoSp}
\begin{array}{rcl}
\cov\big(\epsilon_{Hp}(\bs_i),\epsilon_{Hq}(\bs_j)\big)&=&{\tau\inv_H}v_{pq},\\
\cov\big(\epsilon_{Fp}(\bs_i),\epsilon_{Fq}(\bs_j)\big)&=&{\tau\inv_F}v_{pq}.\\
\end{array}
\end{equation}

\begin{figure}[h!]
	\centering
	\includegraphics[width=\linewidth]{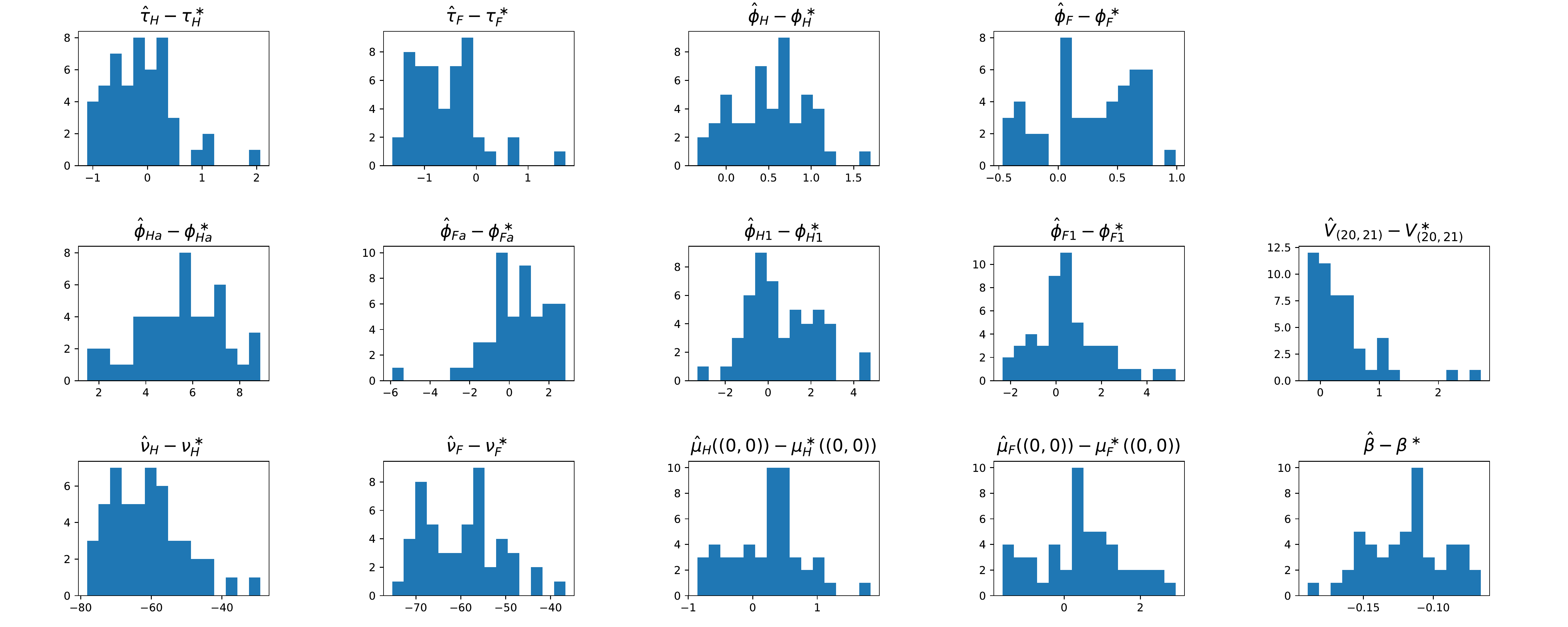}
	\caption{The histograms of the differences between the posterior estimates of parameters or latent states and the true values in the 50 independent experiments where we use Formula~\eqref{eq:NoR} for climate model runs and Formula~\eqref{eq:NoSp} for the random process covariance in the climate model means. The same notation as in Figure~\ref{fig:NoV-Multiple-Runs} is used.}
	\label{fig:NoSp-Multiple-Runs}
\end{figure}

\begin{figure}[h!]
	\centering
	\includegraphics[width=\linewidth]{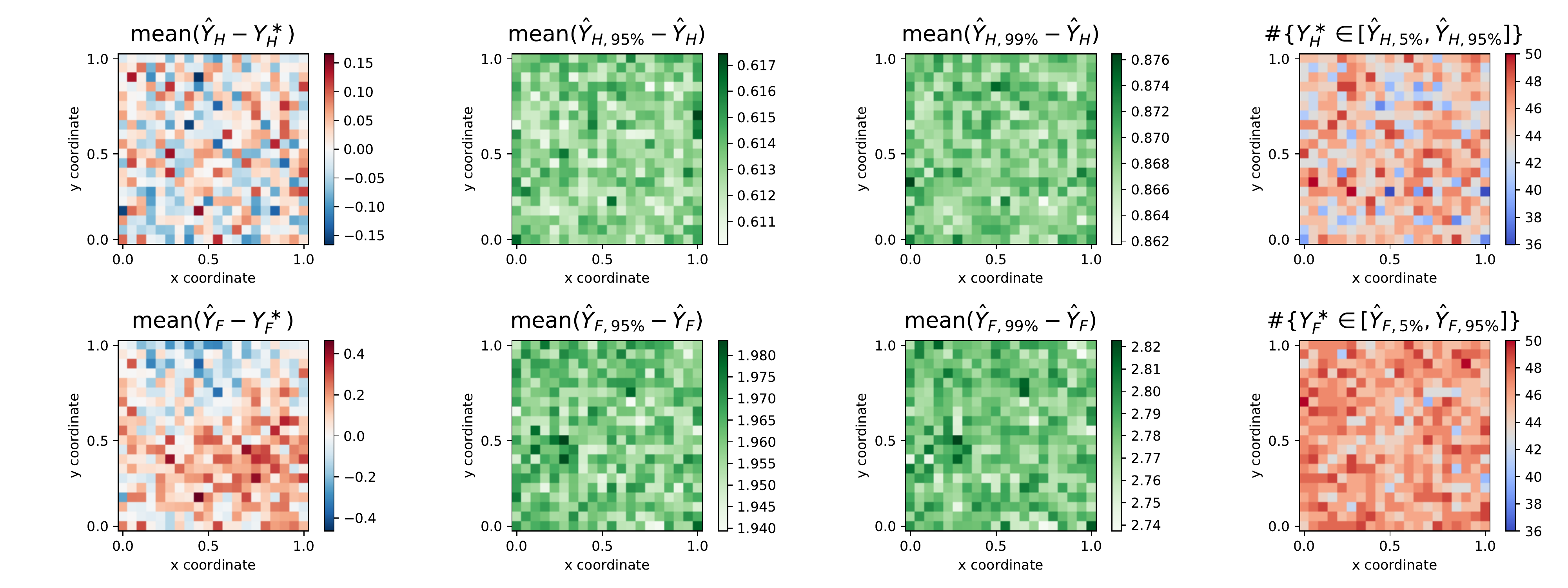}
	\caption{Summary of the expected climate results where we use Formula~\eqref{eq:NoR} for climate model runs and Formula~\eqref{eq:NoSp} for the random process covariance in the climate model means. The same notation as in Figure~\ref{fig:NoV-Field-Multiple-Runs} is used.}
	\label{fig:NoSp-Field-Multiple-Runs}
\end{figure}

We use the same 50 sets of synthetic data as in Section~\ref{sec:simulation}. The posterior inference results are given in  Figures~\ref{fig:NoSp-Multiple-Runs} and \ref{fig:NoSp-Field-Multiple-Runs}. Note that compared to Figure~\ref{fig:NoV-Field-Multiple-Runs}, we use different scales and find larger errors on the posterior means. The estimated variability becomes larger than the full model inference results in Figure~\ref{fig:Full-Field-Multiple-Runs} by looking at the deviations of posterior $95\%$- and $99\%$-quantiles from the posterior means. We also find that the phenomenon for the larger estimate variability in the border than the interior area has disappeared in this case because no spatial correlation is assumed.

\clearpage
\subsubsection{Simplest model inference}\label{sup:simp}
Section~\ref{sup:simp} acts like a combination of Section~\ref{sup:NoV} and Section~\ref{sup:NoSp}, where we ignore both the spatial correlation and the climate model dependence in the climate model means and the spatial correlation in the climate model runs. Therefore, the climate model runs described in Formula~\eqref{eq:modelRuns} are changed to Formula~\eqref{eq:NoR}, and the covariance specified in Formula~\eqref{eq:covForClimateModels} is changed to Formula~\eqref{eq:simp}.
\begin{equation}\label{eq:simp}
\cov\big(\epsilon_{Hp}(\bs_i),\epsilon_{Hq}(\bs_j)\big)={\tau\inv_H},\quad \cov\big(\epsilon_{Fp}(\bs_i),\epsilon_{Fq}(\bs_j)\big)={\tau\inv_F}.
\end{equation}

\begin{figure}[h!]
	\centering
	\includegraphics[width=\linewidth]{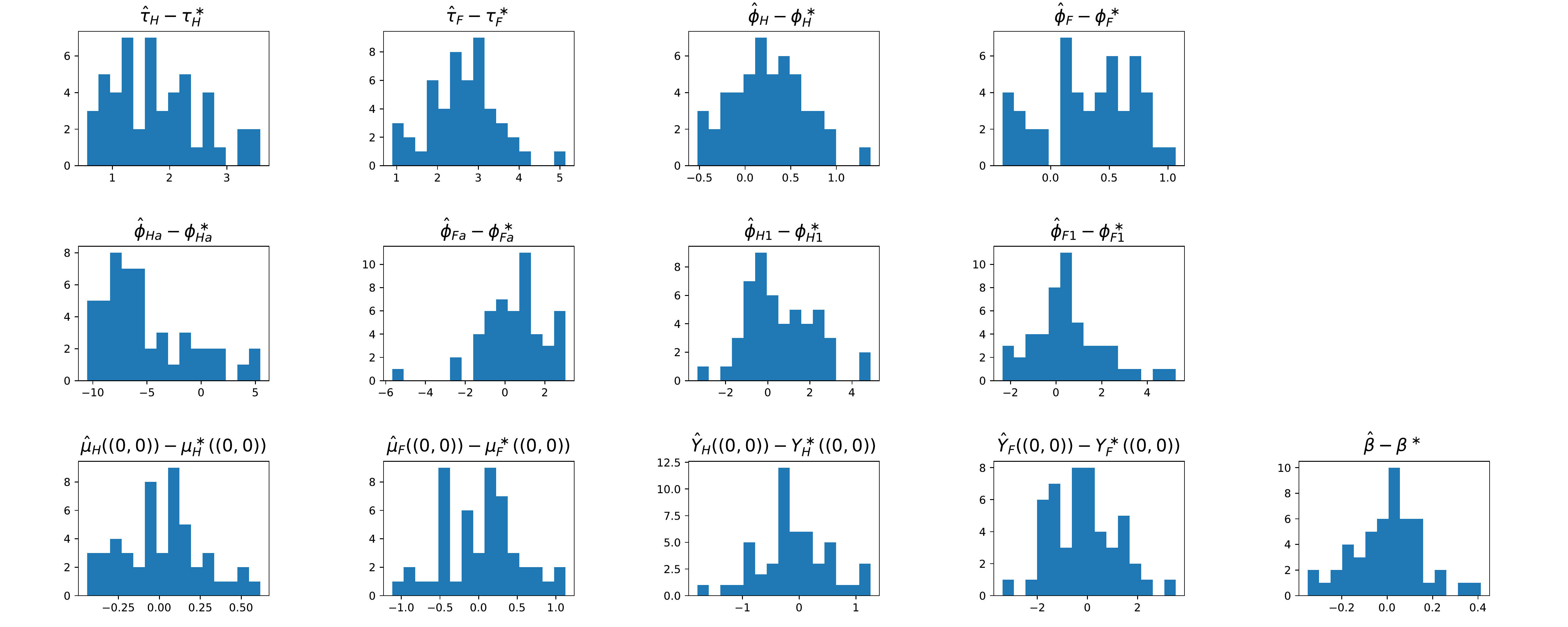}
	\caption{The histograms of the differences between the posterior estimates of parameters or latent states and the true values in the 50 independent experiments where we use Formula~\eqref{eq:NoR} for climate model runs and Formula~\eqref{eq:simp} for the random process covariance in the climate model means. The same notation as in Figure~\ref{fig:NoV-Multiple-Runs} is used.}
	\label{fig:SSB-Multiple-Runs}
\end{figure}

\begin{figure}[h!]
	\centering
	\includegraphics[width=\linewidth]{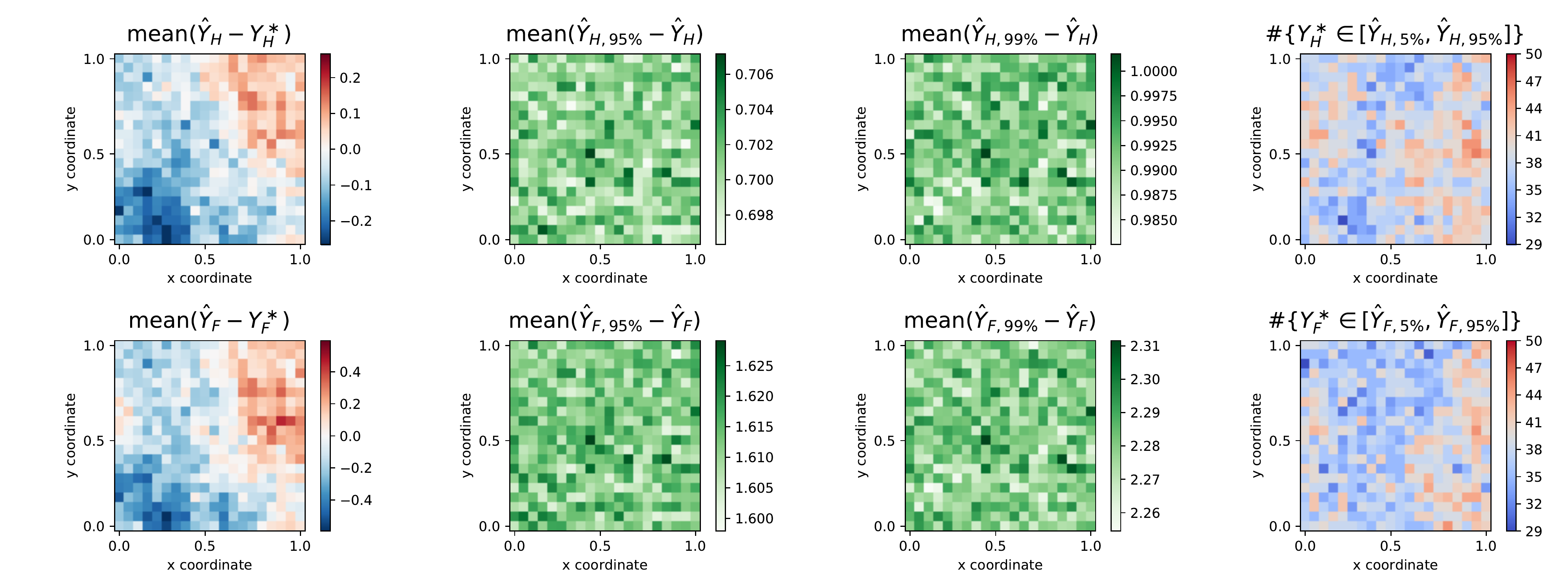}
	\caption{Summary of the expected climate results where we use Formula~\eqref{eq:NoR} for climate model runs and Formula~\eqref{eq:simp} for the random process covariance in the climate model means. The same notation as in Figure~\ref{fig:NoV-Field-Multiple-Runs} is used.}
	\label{fig:SSB-Field-Multiple-Runs}
\end{figure}

Then, this simplest model is similar to that in \cite{Sansom2017} where no spatial random process is used and no climate model dependence is considered. We use the same 50 sets of synthetic data as in Section~\ref{sec:simulation}. The posterior inference results are given in  Figures~\ref{fig:SSB-Multiple-Runs} and \ref{fig:SSB-Field-Multiple-Runs}. We see the underestimated posterior variabilities are similar to those in Section~\ref{sup:NoV}, and the posterior means have much larger errors than all the results presented before. A detailed comparison among all these simplified models and the full model inference results are given in Section~\ref{sup:summary}.

\subsubsection{Summary of the different simplification models in the inference}\label{sup:summary}
\begin{figure}[h!]
	\centering
	\includegraphics[width=\linewidth]{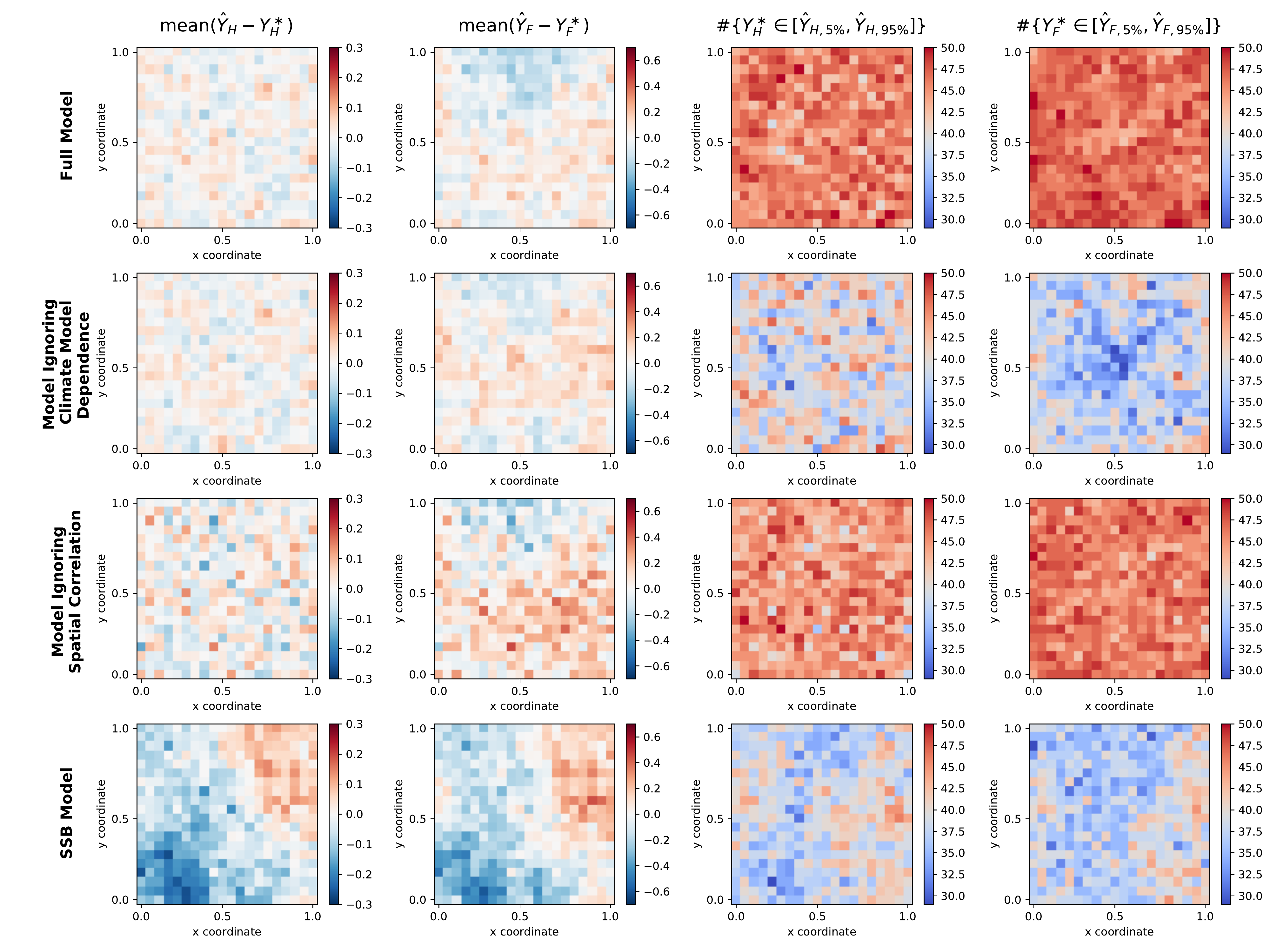}
	\caption{The differences between the sample mean of posterior means ($\hat Y_H$, $\hat Y_F$)  in the 50 independent experiments and the true values ($Y_H^\ast$, $Y_F^\ast$) as well as the number of cases out of the 50 experiments for the true values falling into the $90\%$ credible intervals in different models.}
	\label{fig:Supplimentary-Summary}
\end{figure}

Figure~\ref{fig:Supplimentary-Summary} summarizes the posterior estimate bias of $Y_H$ and $Y_F$ in different model inferences as well as the number of cases out of the 50 experiments for the true parameter values falling into the $90\%$ credible intervals. For ease of comparison, we use the same scale among different model results in Figure~\ref{fig:Supplimentary-Summary}. We see these simplified models lead to larger bias or erroneous variability, and the simplest model (SSB model) shows the largest estimate error in both the mean and the variability.

\clearpage
\subsection{Inference results for synthetic data with the same number of climate model runs and observations as CMIP5}\label{sup:reality}
\begin{figure}[h!]
	\centering
	
	\includegraphics[width=\linewidth]{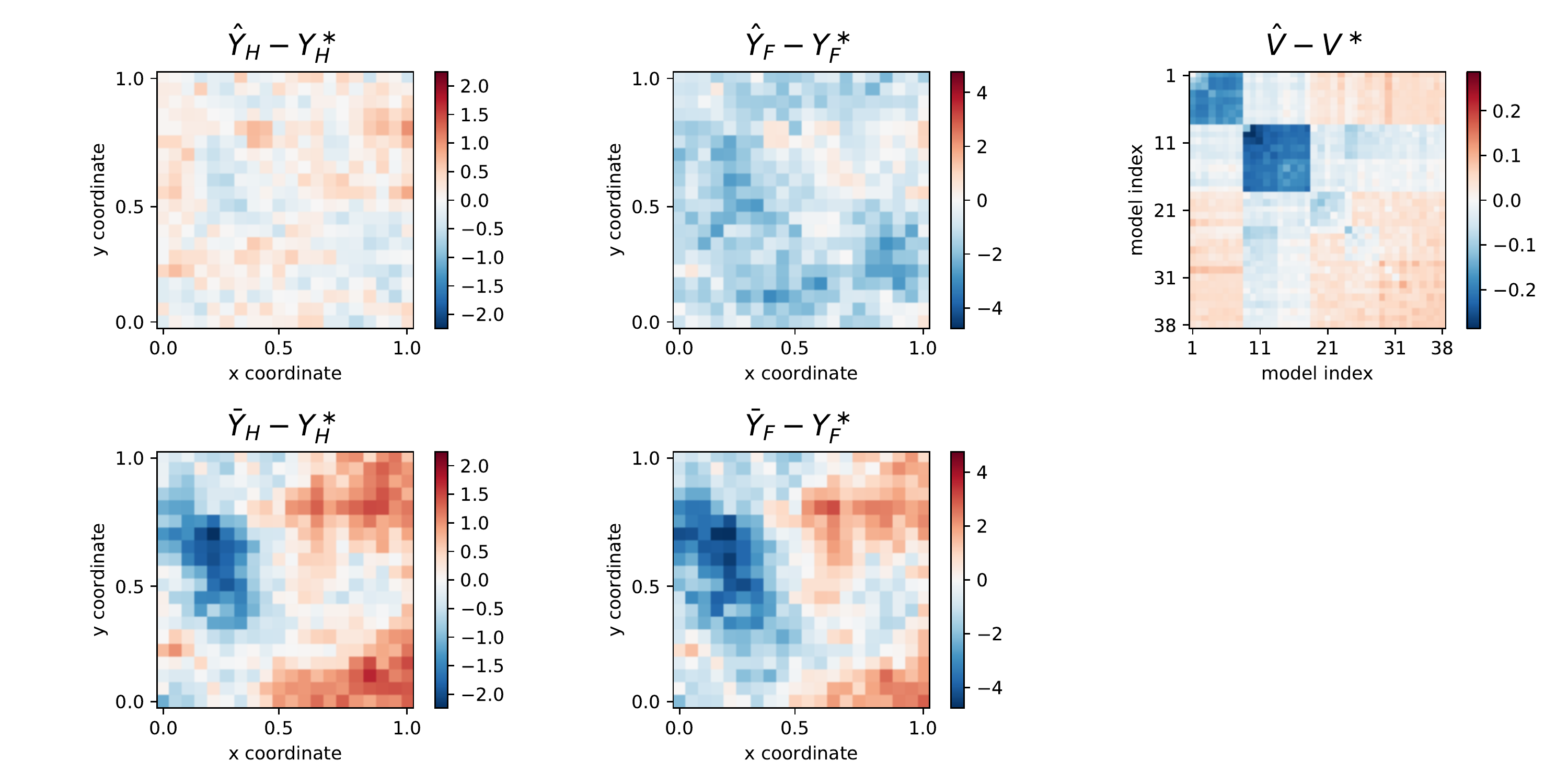}
	\caption{Differences between the estimates and the true values of $Y_H$, $Y_F$, and $V$. The marker $\hat\ $ represents the posterior means in the MCMC using our proposed Bayesian hierarchical model; the marker $\bar\ $ represents the multi-model means calculated as the averages of all the climate model runs; the superscript $^\ast$ represents the true values. The results come from the synthetic data with the same number of climate model runs as in CMIP5 and observations as in the reanalysis products.}
	
	\label{fig:Reality-Field}
\end{figure}

\begin{figure}
	\includegraphics[width=\linewidth]{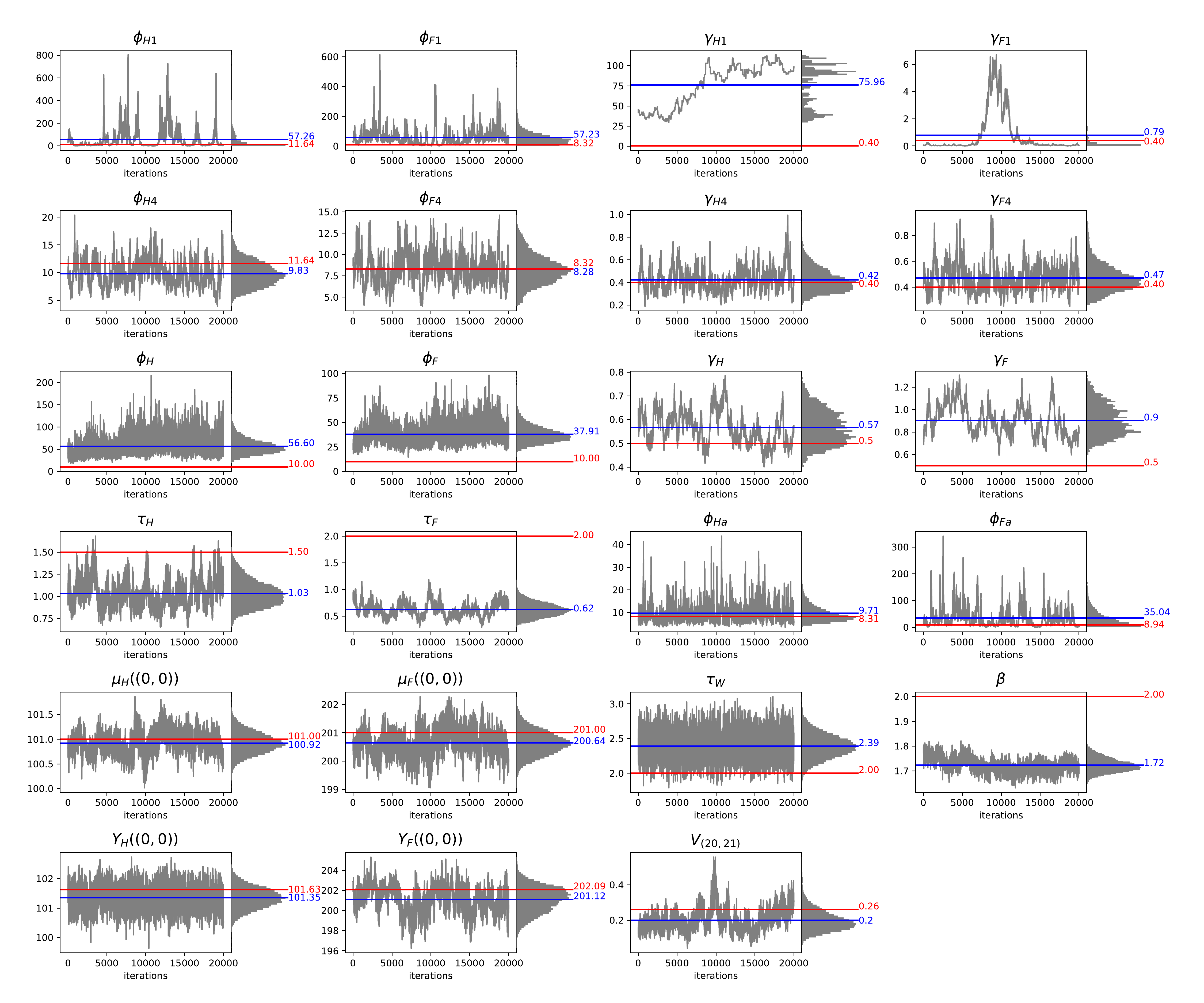}
	\caption{The trace plots, the histograms, the posterior means (blue), and the true values (red) of different parameters or latent states in the MCMC, where we use the synthetic data with the same number of climate model runs as in CMIP5 and observations as in the reanalysis products.}
	
	\label{fig:Reality-Trace}
\end{figure}

\begin{figure}[h!]
	\centering
	\includegraphics[width=\linewidth]{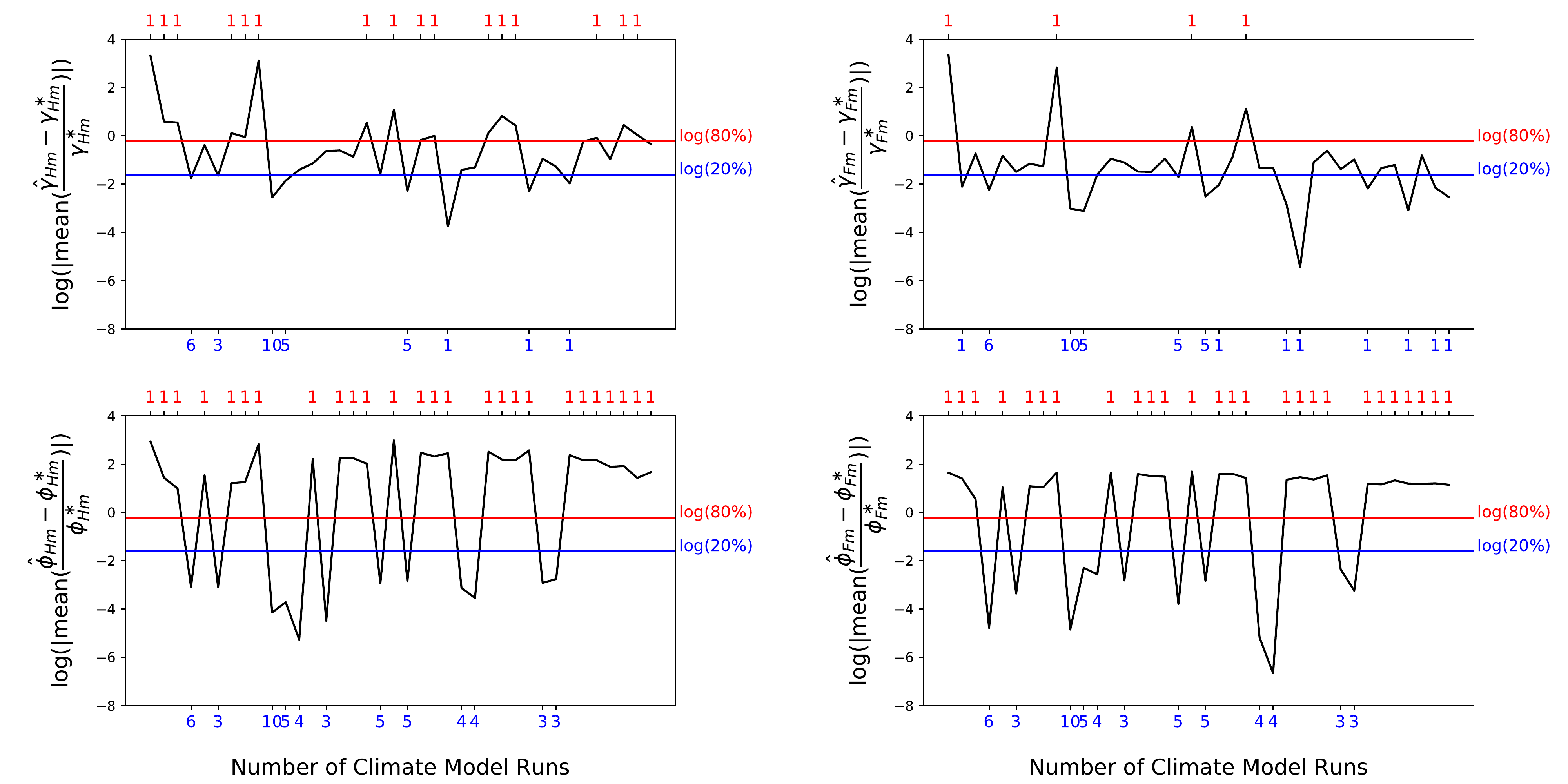}
	\caption{The absolute sample means of the relative differences between the posterior estimates and the true values for $\phi_{Hm},\phi_{Fm},\gamma_{Hm},\gamma_{Fm}$ in the logarithm scale. Two thresholds for $20\%$ and $80\%$ are drawn in each case, respectively. All the climate models above the $80\%$ threshold are marked with the corresponding number of climate model runs on the top; all the climate models below the $20\%$ threshold are marked with the corresponding number of climate model runs on the bottom. The marker $\hat\ $ represents the posterior mean, and the superscript $^\ast$ represents the true values. We use the synthetic data with the same number of climate model runs as in CMIP5 and observations as in the reanalysis products.}
	
	\label{fig:Reality-Spatial}
\end{figure}

\begin{figure}[h!]	
	\includegraphics[width=\linewidth]{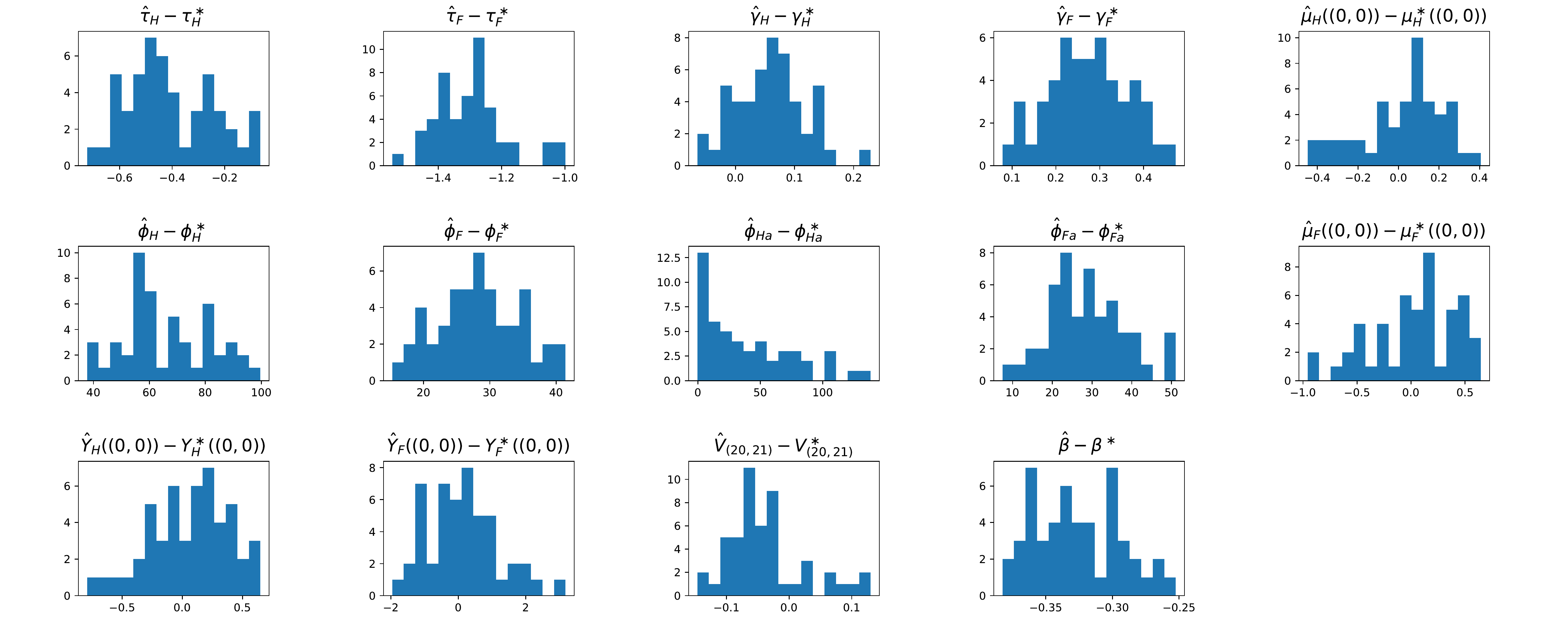}
	\caption{The histogram of the differences between the posterior estimates of parameters or latent states and the true values in the 50 independent experiments. The results come from the inference using the synthetic data with the same number of climate model runs as in CMIP5 and observations as in the reanalysis products.}
	\label{fig:Reality-Multiple-Runs}
\end{figure}

\begin{figure}[h!]
	\centering	
	\includegraphics[width=\linewidth]{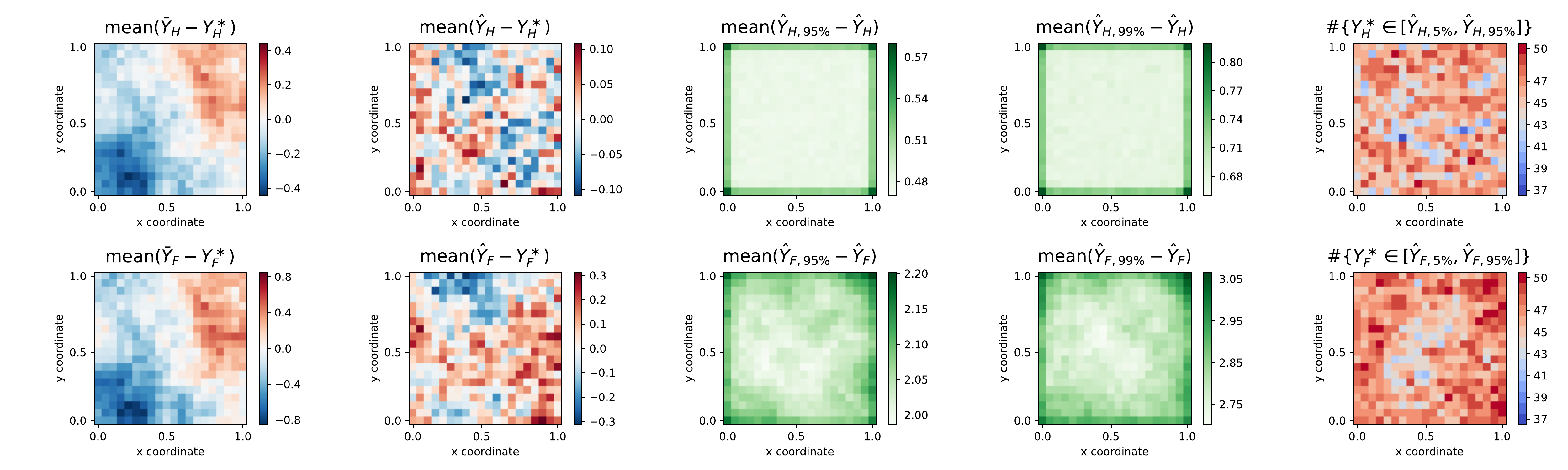}
	\caption{Summary of the results using the synthetic data with the same number of climate model runs as in CMIP5 and observations as in the reanalysis products. The first column shows the sample mean of the differences between the multi-model mean estimates ($\bar Y_H$, $\bar Y_F$) and the true values ($Y_H^\ast$, $Y_F^\ast$) in the 50 independent experiments. The second columns shows the sample mean of the differences between the posterior means ($\hat Y_H$, $\hat Y_F$)  and the true values. The third and forth columns show the sample mean of the differences between the posterior $95\%$-quantiles ($\hat Y_{H,95\%}$, $\hat Y_{F,95\%}$) or the posterior $99\%$-quantiles ($\hat Y_{H,99\%}$, $\hat Y_{F,99\%}$) and the posterior means. The fifth column shows the number of cases out of the 50 experiments for the true values falling into the $90\%$ credible intervals.}
	
	\label{fig:Reality-Field-Multiple-Runs}
\end{figure}

In Section~\ref{sup:reality}, we do not ignore any parts of the model in the inference procedure but reduce the number of climate model runs and observations to make the number of data sets consistent with the available data in CMIP5 and the reanalysis products. More precisely, the number of climate model runs in different climate models are reduced according to Table~\ref{tab:CMIP5}. The number of observation sets is reduced to two, the same as what we have in Section~\ref{sec:application}. Note that we still use the same 50 sets of synthetic data in Section~\ref{sec:simulation} but choose only a subset. The differences between our estimate and the true values of the expected climate and the climate model dependence $V$  are given in Figure~\ref{fig:Reality-Field}. 
For comparison, we also show the difference between the multi-model mean results and the true values in Figure~\ref{fig:Reality-Field}. Trace plots and histograms in the MCMC in one experiment are given in Figure~\ref{fig:Reality-Trace}. 
We see though the expected climate $Y_H$ and $Y_F$ still have good posterior estimates and perform much better than the multi-model mean results,
some other parameters have biased posterior means or do not reach stationary distributions in the MCMC.
In addition, looking at the spatial parameters $\phi_{Hm},\phi_{Fm},\gamma_{Hm},\gamma_{Fm}$ for climate model 1, which has one model run, and climate model 4, which has 6 model runs, illustrated in the first two rows in Figure~\ref{fig:Reality-Trace}, we see the posterior estimation of spatial parameters heavily depends on the number of climate model runs, where $\phi_{H4},\phi_{F4},\gamma_{H4},\gamma_{F4}$ have much better posterior distributions and we do not really get good stationary distribution for $\phi_{H1},\phi_{F1},\gamma_{H1},\gamma_{F1}$. To have a better view on this, we summarize the spatial parameter posterior estimates associated with different climate models in the 50 experiments in Figure~\ref{fig:Reality-Spatial},
where the absolute sample means of the relative differences between the posterior estimates and the true values are given and the climate models that have relative errors less than $20\%$ or greater than $80\%$ are highlighted. We see most of the cases for the accurate posterior estimates are associated with climate models with more than one model runs and all the bad estimators come from the climate models with only one model run.
The histograms of other parameter posterior means in the 50 experiments are given in Figures~\ref{fig:Reality-Multiple-Runs} and the posterior details about the expected climate are given in Figure~\ref{fig:Reality-Field-Multiple-Runs}. Compared to what we get in Section~\ref{sec:simulation}, we see that the estimates of $Y_H$ and $Y_F$ still have similar posterior means but slightly larger parameter estimate variabilities and longer credible intervals. However, $\beta$ is more underestimated.

In conclusion, given the data with the same size as CMIP5, we are less confident about the estimate of the emergent relationship. However, the estimate for the future expected climate $Y_F$ is still accurate and can show us what the future may look like under given forcings by use of the climate model outputs. Furthermore, the estimates of the spatial parameters in the proposed Bayesian hierarchical model are mostly accurate for the climate models with more than one model runs. On the contrary, we generally do not obtain satisfactory posterior results for spatial parameters associated with climate models with only one run. If the emergent relationship and spatial parameters are also of great interest, more runs of climate models are required, especially for those with only one run.

\clearpage
\section{More CMIP5 data details}\label{sup:data}

Table~\ref{tab:CMIP5} gives the number of model runs available for each model in CMIP5.

\begin{table}[htbp]
	\centering
	\caption{Name and the corresponding number of model runs for each climate model in CMIP5. ``R'' stands for the number of model runs. ``I'' stands for the arbitrarily assigned model index in our study when we present our application results.}
	\begin{tabular}{ccc|ccc|ccc}
		\hline
		Model Name & R & I & Model Name & R & I & Model Name & R & I  \\
		\hline
ACCESS1-0	&	1	&	1	&	FIO-ESM	&	3	&	14	&	IPSL-CM5B-LR	&	1	&	27	\\
ACCESS1-3	&	1	&	2	&	GFDL-CM3	&	1	&	15	&	MIROC-ESM	&	1	&	28	\\
BNU-ESM	&	1	&	3	&	GFDL-ESM2G	&	1	&	16	&	MIROC-ESM-CHEM	&	1	&	29	\\
CCSM4	&	6	&	4	&	GFDL-ESM2M	&	1	&	17	&	MIROC5	&	3	&	30	\\
CESM1-BGC	&	1	&	5	&	GISS-E2-H	&	5	&	18	&	MPI-ESM-LR	&	3	&	31	\\
CESM1-CAM5	&	3	&	6	&	GISS-E2-H-CC	&	1	&	19	&	MPI-ESM-MR	&	1	&	32	\\
CMCC-CM	&	1	&	7	&	GISS-E2-R	&	5	&	20	&	MRI-CGCM3	&	1	&	33	\\
CMCC-CMS	&	1	&	8	&	GISS-E2-R-CC	&	1	&	21	&	NorESM1-M	&	1	&	34	\\
CNRM-CM5	&	1	&	9	&	HadGEM2-AO	&	1	&	22	&	NorESM1-ME	&	1	&	35	\\
CSIRO-Mk3-6-0	&	10	&	10	&	HadGEM2-CC	&	1	&	23	&	bcc-csm1-1	&	1	&	36	\\
CanESM2	&	5	&	11	&	HadGEM2-ES	&	4	&	24	&	bcc-csm1-1-m	&	1	&	37	\\
EC-EARTH	&	4	&	12	&	IPSL-CM5A-LR	&	4	&	25	&	inmcm4	&	1	&	38	\\
FGOALS-g2	&	1	&	13	&	IPSL-CM5A-MR	&	1	&	26	&		&		&		\\
		\hline
	\end{tabular}
	\label{tab:CMIP5}
\end{table}

\section{More MCMC results for the application}\label{sup:app}
\begin{figure}[t!]
	\centering	
	
	\includegraphics[width=\linewidth]{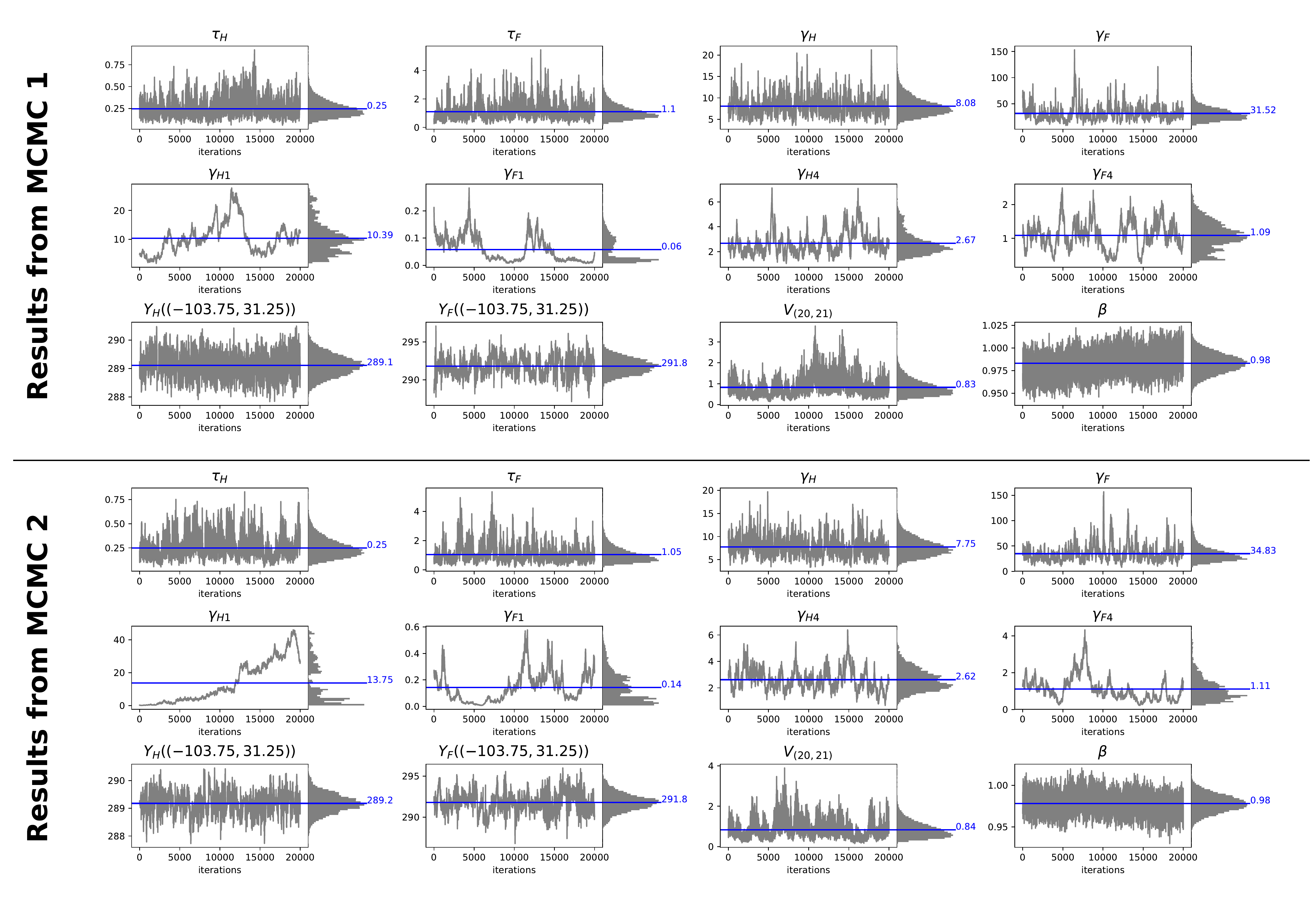}
	\caption{Posterior distributions and posterior means of different parameters and latent states in two independent MCMC runs with our proposed Bayesian hierarchical model using the climate model outputs under RCP4.5 for the future forecast in the CNA region.}
	\label{fig:CNA-Trace}
\end{figure}

Figure~\ref{fig:CNA-Trace} gives two trace plot results of selected parameters and latent states from two MCMC runs for our proposed Bayesian hierarchical model using climate model outputs under RCP4.5 for the future forecast in the CNA region.
We see that in different MCMC runs, there is a certain variation in the posterior estimates of spatial parameters possibly due to the less stable distribution of $\gamma_{Hm}$ and $\gamma_{Fm}$, especially for those associated with climate models that have only one model run (for example, $\gamma_{H1}$ and $\gamma_{F1}$). However, the expected climate $Y_H,Y_F$, which we are mostly interested in, and the emergent relationship $\beta$ have quite stable posterior means. In Section~\ref{sup:reality} where the number of climate model runs and observations are reduced to the same as CMIP5, we see that $\gamma_{Hm}$ and $\gamma_{Fm}$ estimates are quite noisy, but the estimate  $Y_H$ and $Y_F$ as well as their uncertainties are still accurate. The estimate of $\beta$ tends to be underestimated. Hence, we are confident about the accuracy of the posterior results of $Y_H$ and $Y_F$ and think that $\beta$ may be a little underestimated. We also believe that if some climate models (especially for those that only have one model run) can be run more times and provide more replicates, the estimates of $\gamma_{Hm}$ and $\gamma_{Fm}$ will be much better and the underestimation of $\beta$ will be alleviated.

\section{Formulae for the Gibbs updates}\label{sup:formulae}
We introduce some notation for aggregated variables for ease of presentation.  Assume that we have data over $n$ locations $\bs_1,\ldots,\bs_n\in\mathcal{D}$. We denote the climate model outputs by
$\mathbf{X}=\big(X_{Hmr}(\bs),X_{Fmr^\prime}(\bs):m=1,\ldots,M,r=1,\ldots,R_{Hm},r^\prime=1,\ldots,R_{Fm},\bs=\bs_1,\ldots,\bs_n\big)$. We also denote all the observations by
$\mathbf{W}=\big(W_i(\bs):\bs=\bs_1,\ldots,\bs_n, i=1,\ldots,N\big).$
For the latent states, we denote 
$
\boldsymbol{\chi}=(X_{Hm}(\bs),X_{Fm}(\bs):m=1,\ldots,M,\bs=\bs_1,\ldots,\bs_n),$ 
$
\mathbf{Y}=(Y_H(\bs),Y_F(\bs),Y_{Ha}(\bs),Y_{Fa}(\bs):\bs=\bs_1,\ldots,\bs_n),
$
and $\boldsymbol{\xi}=(\mu_H(\bs),\mu_F(\bs):\bs=\bs_1,\ldots,\bs_n).$
For the parameters, we denote
$
\blambda=(\phi_{Hm},\phi_{Fm},\gamma_{Hm}, \gamma_{Fm}: m=1,\ldots,M)
$ as a vector of parameters involved in the spatial models of the climate model mean and 
$
\btheta=(\beta,\tau_W,\phi_{Ha},$ $\phi_{Fa},\phi_H,\phi_F,\nu_H,
\nu_F,\tau_H,\tau_F,\gamma_H,\gamma_F,V)
$ the vector of all other parameters.

The joint posterior of $\bchi,\bY,\bxi,\blambda,\btheta$ given $\bX, \bW$ is
\[
\scriptsize
\begin{array}{ll}
&\Pr(\bchi,\bY,\bxi,\blambda,\btheta\mid \bX,\bW)\\
\propto~&\Pr( \bX,\bW\mid\bchi,\bY,\bxi,\blambda,\btheta)\Pr(\bchi,\bY,\bxi,\blambda,\btheta)\\
=~&
\Pr(\bX\mid\bchi,\blambda)
\Pr(\bW\mid \bY,\btheta)
\Pr(\bchi\mid \btheta,\bxi)
\Pr(\bY\mid \bxi,\btheta)
\Pr(\bxi)
\Pr(\blambda\mid \btheta)
\Pr(\btheta).
\end{array}
\]

We also define some vectors and matrices as follows to present the formulae in this section.
	\[
	\scriptsize
	\begin{array}{llllll}
	\by_H&=&\big(Y_H(\bs_1),\ldots,Y_H(\bs_n)\big)\trans,&\quad
	\by_F&=&\big(Y_F(\bs_1),\ldots,Y_F(\bs_n)\big)\trans,\\
	\by_{Ha}&=&\big(Y_{Ha}(\bs_1),\ldots,Y_{Ha}(\bs_n)\big)\trans,&\quad
	\by_{Fa}&=&\big(Y_{Fa}(\bs_1),\ldots,Y_{Fa}(\bs_n)\big)\trans,\\
	\bx_{H}(\bs)&=&\big(X_{H1}(\bs),\ldots,X_{HM}(\bs)\big)\trans,&\quad
	\bx_{F}(\bs)&=&\big(X_{F1}(\bs),\ldots,X_{FM}(\bs)\big)\trans,\\
	X_H&=&\big(\bx_H(\bs_1),\ldots,\bx_H(\bs_n)\big)\trans,&\quad
	X_F&=&\big(\bx_F(\bs_1),\ldots,\bx_F(\bs_n)\big)\trans.\\
	\bx_{Hm}&=&\big(X_{Hm}(\bs_1),\ldots,X_{Hm}(\bs_n)\big)\trans,&\quad
	\bx_{Fm}&=&\big(X_{Fm}(\bs_1),\ldots,X_{Fm}(\bs_n)\big)\trans,\\
	\bx_{Hmr}&=&\big(X_{Hmr}(\bs_1),\ldots,X_{Hmr}(\bs_n)\big)\trans,&\quad
	\bx_{Fmr}&=&\big(X_{Fmr}(\bs_1),\ldots,X_{Fmr}(\bs_n)\big)\trans,\\
	\bmu_{H}&=&\big(\bmu_H(\bs_1),\ldots,\bmu_H(\bs_n)\big)\trans,&\quad
	\bmu_{F}&=&\big(\bmu_F(\bs_1),\ldots,\bmu_F(\bs_n)\big)\trans.\\
	\bw_i&=&\big(W_i(\bs_1),\ldots,W_i(\bs_n)\big)\trans.&&&\\
	\end{array}
	\]
	
	The probability density function of $\bY$ given $\bW,\btheta,\bxi$ is,
	\[
	\scriptsize
	\begin{array}{rl}
	&\Pr(\bY\mid \bW,\btheta,\bxi)\\
	\propto&\Pr(\bW\mid \bY,\btheta)\Pr(\bY\mid \btheta,\bxi)\\
	\propto
	&\displaystyle\prod_{i=1}^{N}\exp\left(-\frac{\tau_W}{2}(\by_{Ha}-\bw_i)\trans(\by_{Ha}-\bw_i)\right)\times\\
	&\exp\left(-\dfrac{\tau_H}{2\kappa}(\by_H-\bmu_H)\trans\Sigma\inv_H(\by_H-\bmu_H)\right)\times\\
	&\exp\left(-\dfrac{\tau_F}{2\kappa}\big\{\by_F-\bmu_F-\beta(\by_H-\bmu_H)\big\}\trans\Sigma_F\inv\big\{\by_F-\bmu_F-\beta(\by_H-\bmu_H)\big\}\right)\times\\
	&\exp\left(-\dfrac{\phi_{Ha}}{2}(\by_{Ha}-\by_H)\trans(\by_{Ha}-\by_H)\right)\times
	\exp\left(-\dfrac{\phi_{Fa}}{2}(\by_{Fa}-\by_F)\trans(\by_{Fa}-\by_F)\right),\\
	\end{array}
	\]
	where ${\Sigma_H}$ is an $n\times n$ matrix with $(i,j)$-th entry as $c(\|\bs_i-\bs_j\|;\gamma_H)$, $\Sigma_F$ is an $n\times n$ matrix with $(i,j)$-th entry as $c(\|\bs_i-\bs_j\|;\gamma_F)$, and $c(\cdot;\cdot)$ is the Whittle covariance function.
	Then, the full conditional distribution of $\bY$ is,
	\[
	\scriptsize
	\begin{array}{rcl}
	\by_{Fa}\mid\ldots&\sim&N\big(\by_F,\phi_{Fa}\inv I\big),\\
	\by_{Ha}\mid\ldots&\sim&N\bigg(\dfrac{1}{\phi_{Ha}+N\tau_W}(\phi_{Ha}\by_{H}+\displaystyle\sum_{i=1}^{N}\tau_W\bw_i),(\phi_{Ha}+N\tau_W)\inv I\bigg),\\
	\by_{F}\mid\ldots&\sim&N\bigg(\big(\dfrac{\tau_F}{\kappa}\Sigma_F\inv+\phi_{Fa}I\big)\inv\big[\dfrac{\tau_F}{\kappa}\Sigma_F\inv\big\{\bmu_F+\beta(\by_H-\bmu_H)\big\}+\phi_{Fa}\by_{Fa}\big],\\
	&&~~~~~\big(\dfrac{\tau_F}{\kappa}\Sigma_F\inv+\phi_{Fa}I\big)\inv\bigg),\\
	\by_{H}\mid\ldots&\sim&N\bigg(\big(\dfrac{\tau_F\beta^2}{\kappa}\Sigma_F\inv+\dfrac{\tau_H}{\kappa}\Sigma_H\inv+\phi_{Ha}I\big)\inv\big(\dfrac{\tau_F\beta}{\kappa}\Sigma_F\inv(\by_F-\bmu_F+\beta\bmu_H)+\dfrac{\tau_H}{\kappa}\Sigma_H\inv\bmu_H+\phi_{Ha}\by_{Ha}\big),\\
	&&~~~~~\big(\dfrac{\tau_F\beta^2}{\kappa}\Sigma_F\inv+\dfrac{\tau_H}{\kappa}\Sigma_H\inv+\phi_{Ha}I\big)\inv\bigg),\\
	\end{array}
	\]
	where $I$ is the $n\times n$ identity matrix.
	
	The probability density function of $\bchi$ given $\bX,\btheta,\bxi,\blambda$ is
	\[
	\scriptsize
	\begin{array}{rl}
	&\Pr(\bchi\mid \bX,\btheta,\bxi,\blambda)\\
	\propto&\Pr(\bX\mid \bchi,\blambda)\Pr(\bchi\mid \btheta,\bxi)\\
	\propto
	&\displaystyle\prod_{m=1}^{M}\displaystyle\prod_{r=1}^{R_{Hm}} \exp\left(-\frac{\phi_{Hm}}{2}(\bx_{Hmr}-\bx_{Hm})\trans\Sigma_{Hm}\inv(\bx_{Hmr}-\bx_{Hm})\right)\times\\
	&\displaystyle\prod_{m=1}^{M}\displaystyle\prod_{r=1}^{R_{Fm}} \exp\left(-\frac{\phi_{Fm}}{2}(\bx_{Fmr}-\bx_{Fm})\trans\Sigma_{Fm}\inv(\bx_{Fmr}-\bx_{Fm})\right)\times\\
	&\exp\left(-\dfrac{\tau_H}{2}  \big\{\vecop(X_{Hm})-\bone_M\otimes\bmu_H\big\}\trans (V\otimes \Sigma_H)\inv \big\{\vecop(X_{Hm})-\bone_M\otimes\bmu_H\big\}\right)\times\\
	&\exp\left(-\dfrac{\tau_F}{2} \big[\vecop(X_{Fm})-\bone_M\otimes\bmu_F-\beta\big\{\vecop(X_{Hm})-\bone_M\otimes\bmu_H\big\}\big]\trans (V\otimes \Sigma_F)\inv \times\right.\\
	&\big[\vecop(X_{Fm})-\bone_M\otimes\bmu_F-\beta\big\{\vecop(X_{Hm})-\bone_M\otimes\bmu_H\big\}\big]\bigg),\\
	\end{array}
	\]
	where $\bone_M$ is the $M$-dimensional vector with all values equal to one,  $\otimes$ is the Kronecker product, ${\Sigma_{Hm}}$ is an $n\times n$ matrix with $(i,j)$-th entry as $c(\|\bs_i-\bs_j\|;\gamma_{Hm})$, $\Sigma_{Fm}$ is an $n\times n$ matrix with $(i,j)$-th entry as $c(\|\bs_i-\bs_j\|;\gamma_{Fm})$, and $c(\cdot;\cdot)$ is the Whittle covariance function.
	
	When $Mn$ (the dimension of the matrix $V\otimes \Sigma_H$) is large, it is computational expensive to update all $\bx_{Hm}$ at the same time, so we choose to update each $\bx_{Hm}$ sequentially. This also applies for $\bx_{Fm}$. We define some notation for operations to an arbitrary matrix $K$. $K_{(m,m)}$ denotes the element in the $m$-th row and $m$-th column of $K$, $K_{(i:j,m)}$ denotes the trimmed $m$-th column of $K$ consisting of values from $i$-th row to $j$-th row, $K_{(m,i:j)}$ denotes the trimmed $m$-th row of $K$ consisting of values from $i$-th column to $j$-th column, $K_{(i:j,i:j)}$ denotes the sub-matrix of $K$ consisting of values from $i$-th row to $j$-th row and from $i$-th column to $j$-th column. Then, the full conditional distribution of $\bchi$ is,
	\[
	\scriptsize
	\begin{array}{rcl}
	\bx_{Fm}\mid\ldots&\sim&N\bigg((R_{Fm}\phi_{Fm}\Sigma_{Fm}\inv+Q_F\inv)\inv(\phi_{Fm}\Sigma_{Fm}\inv\sum^{R_{Fm}}_{r=1}\bx_{Fmr}+Q_F\inv\boldsymbol{v}_F),(R_{Fm}\phi_{Fm}\Sigma_{Fm}\inv+Q_F\inv)\inv\bigg),\\
	\bx_{Hm}\mid\ldots&\sim&N\bigg((R_{Hm}\phi_{Hm}\Sigma_{Hm}\inv+Q_H\inv+\beta^2Q_F\inv)\inv(\phi_{Hm}\Sigma_{Hm}\inv\sum^{R_{Hm}}_{r=1}\bx_{Hmr}+Q_H\inv\boldsymbol{v}_H+Q_F\inv\boldsymbol{v}'_H),\\
	&&~~~~~(R_{Hm}\phi_{Hm}\Sigma_{Hm}\inv+Q_H\inv+\beta^2Q_F\inv)\inv\bigg),\\
	\end{array}
	\]
	where $I$ is the $n\times n$ identity matrix, and 
	\[
	\scriptsize
	\begin{array}{rcl}
	\boldsymbol{v}_F&=&\bmu_F+\beta(\bx_{Hm}-\bmu_H)+(V_{(m,1:m-1)}V\inv_{(1:m-1,1:m-1)}\otimes I)\times\\
	&&\big[\vecop({X_{Fm}}_{(1:n,1:m-1)})-\bone_{m-1}\otimes\bmu_F-\beta\big\{\vecop({X_{Hm}}_{(1:n,1:m-1)})-\bone_{m-1}\otimes\bmu_H\big\}\big]\\
	&=&\bmu_F+\beta(\bx_{Hm}-\bmu_H)+\\
	&&({X_{Fm}}_{(1:n,1:m-1)}-\beta{X_{Hm}}_{(1:n,1:m-1)})V\inv_{(1:m-1,1:m-1)} V_{(1:m-1,m)}+\\
	&&(V_{(m,1:m-1)}V\inv_{(1:m-1,1:m-1)}\bone_{m-1})(\beta\bmu_H-\bmu_F),\\
	
	Q_F&=&\tau_F\inv\big\{V_{(m,m)}-V_{(m,1:m-1)}V_{(1:m-1,1:m-1)}\inv V_{(1:m-1,m)}\big\}\Sigma_F,\\
	
	\boldsymbol{v}_H&=&\bmu_H+(V_{(m,1:m-1)}V\inv_{(1:m-1,1:m-1)}\otimes I)
	\big\{\vecop({X_{Hm}}_{(1:n,1:m-1)})-\bone_{m-1}\otimes\bmu_H\big\}\\
	&=&\bmu_H+{X_{Hm}}_{(1:n,1:m-1)}V\inv_{(1:m-1,1:m-1)} V_{(1:m-1,m)}-
	(V_{(m,1:m-1)}V\inv_{(1:m-1,1:m-1)}\bone_{m-1})\bmu_H,\\
	
	\boldsymbol{v}'_H&=&
	\beta\big[\bx_{Fm}-\bmu_F+\beta\bmu_H+(V_{(m,1:m-1)}V\inv_{(1:m-1,1:m-1)}\otimes I)\\
	&&
	\big\{\beta\vecop({X_{Hm}}_{(1:n,1:m-1)})-\vecop({X_{Fm}}_{(1:n,1:m-1)})+\bone_{m-1}\otimes\bmu_F-\beta\bone_{m-1}\otimes\bmu_H\big\}\big]\\
	&=&\beta\big\{\bx_{Fm}-\bmu_F+\beta\bmu_H-\\
	&&({X_{Fm}}_{(1:n,1:m-1)}-\beta{X_{Hm}}_{(1:n,1:m-1)})V\inv_{(1:m-1,1:m-1)} V_{(1:m-1,m)}+\\
	&&(V_{(m,1:m-1)}V\inv_{(1:m-1,1:m-1)}\bone_{m-1})(\beta\bmu_H-\bmu_F)\big\},\\
	
	Q_H&=&\tau_H\inv\big\{V_{(m,m)}-V_{(m,1:m-1)}V_{(1:m-1,1:m-1)}\inv V_{(1:m-1,m)}\big\}\Sigma_H.\\
	\end{array}
	\]
	
	The probability density function of $\bxi$ given $\bY,\bchi,\btheta$ is
	\[
	\scriptsize
	\begin{array}{rl}
	&\Pr(\bxi\mid \bY,\bchi,\btheta)\\
	\propto&\Pr(\bY\mid \bxi, \btheta)\Pr(\bchi\mid \bxi, \btheta)\Pr(\bxi)\\
	\propto
	&\exp\left(-\dfrac{\tau_H}{2\kappa}(\by_H-\bmu_H)\trans\Sigma_H\inv(\by_H-\bmu_H)\right)\times\\
	&\exp\left(-\dfrac{\tau_F}{2\kappa}\big\{\by_F-\bmu_F-\beta(\by_H-\bmu_H)\big\}\trans\Sigma_F\inv\big\{\by_F-\bmu_F-\beta(\by_H-\bmu_H)\big\}\right)\times\\
	&\exp\left(-\dfrac{\tau_H}{2}\big\{\vecop(X_{Hm})-\bone_M\otimes\bmu_H\big\}\trans (V\otimes \Sigma_H)\inv \big\{\vecop(X_{Hm})-\bone_M\otimes\bmu_H\big\}\right)\times\\
	&\exp\left(-\dfrac{\tau_F}{2}\big[\vecop(X_{Fm})-\bone_M\otimes\bmu_F-\beta\big\{\vecop(X_{Hm})-\bone_M\otimes\bmu_H\big\}\big]\trans (V\otimes \Sigma_F)\inv \times\right.\\
	&\big[\vecop(X_{Fm})-\bone_M\otimes\bmu_F-\beta\big\{\vecop(X_{Hm})-\bone_M\otimes\bmu_H\big\}\big]\bigg)\times\\
	&\exp\left(-\dfrac{10^{-6}}{2}\bmu_H\trans\bmu_H\right)\exp\left(-\dfrac{10^{-6}}{2}\bmu_F\trans\bmu_F\right),
	\end{array}
	\]	
	
	Then, the full conditional distribution of $\bxi$ is
	\[
	\scriptsize
	\begin{array}{rcl}
	\bmu_F\mid\ldots&\sim&N\bigg(\big(\dfrac{\tau_F}{\kappa}\Sigma_F\inv+Q_{\bmu_F}\inv+10^{-6}I\big)\inv\big[\dfrac{\tau_F}{\kappa}\Sigma_F\inv\big\{\by_F-\beta(\by_H-\bmu_H)\big\}+Q_{\bmu_F}\inv\boldsymbol{v}_{\bmu_F}\big],\\
	&&\big\{\dfrac{\tau_F}{\kappa}\Sigma_F\inv+Q_{\bmu_F}\inv+10^{-6}I\big\}\inv\bigg),\\
	
	\bmu_H\mid\ldots&\sim&N\bigg(\big(\dfrac{\tau_H}{\kappa}\Sigma_H\inv+\dfrac{\tau_F\beta^2}{\kappa}\Sigma_F\inv+
	Q_{\bmu_H}\inv+\beta^2Q_{\bmu_F}\inv+10^{-6}I\big)\inv\times\\
	&&\big\{\dfrac{\tau_H}{\kappa}\Sigma_H\inv\by_H+\dfrac{\tau_F\beta}{\kappa}\Sigma_F\inv\big(\bmu_F+\beta\by_H-\by_F\big)+
	Q_{\bmu_H}\inv\boldsymbol{v}_{\bmu_H}+Q_{\bmu_F}\inv\boldsymbol{v}_{\bmu_H}'\big\},\\
	&&\big(\dfrac{\tau_H}{\kappa}\Sigma_H\inv+\dfrac{\tau_F\beta^2}{\kappa}\Sigma_F\inv+
	Q_{\bmu_H}\inv+\beta^2Q_{\bmu_F}\inv+10^{-6}I\big)\inv\bigg),\\
	\end{array}
	\]
	where
	\[
	\scriptsize
	\begin{array}{rcl}
	\boldsymbol{v}_{\mu_F}&=&\beta\bmu_H+(\bone_M\trans V\inv \bone_M)\inv(X_{Fm}-\beta X_{Hm})V\inv\bone_M,\\
	Q_{\bmu_F}&=&\tau_F\inv(\bone_M\trans V\inv \bone_M)\inv\Sigma_F,\\
	\boldsymbol{v}_{\mu_H}&=&(\bone_M\trans V\inv \bone_M)\inv X_{Hm}V\inv\bone_M,\\
	\boldsymbol{v}_{\mu_H}'&=&\beta\big\{\bmu_F-(\bone_M\trans V\inv \bone_M)\inv(X_{Fm}-\beta X_{Hm})V\inv\bone_M\big\},\\
	Q_{\bmu_H}&=&\tau_H\inv(\bone_M\trans V\inv \bone_M)\inv\Sigma_H.\\
	\end{array}
	\]
	
	The probability of $\blambda$ given $\bX,\bchi,\btheta$ is
	\[
	\scriptsize
	\begin{array}{rl}
	&\Pr(\blambda\mid\bX,\bchi,\btheta)\\
	\propto&\Pr(\bX\mid \bchi, \blambda, \btheta)\Pr(\blambda\mid \btheta)\\
	\propto
	&\displaystyle\prod_{m=1}^{M}\displaystyle\prod_{r=1}^{R_{Hm}} \phi^{n/2}_{Hm}\det(\Sigma_{Hm})^{-1/2}\exp\left(-\frac{\phi_{Hm}}{2}(\bx_{Hmr}-\bx_{Hm})\trans\Sigma_{Hm}\inv(\bx_{Hmr}-\bx_{Hm})\right)\times\\
	&\displaystyle\prod_{m=1}^{M}\displaystyle\prod_{r=1}^{R_{Fm}} \phi^{n/2}_{Fm}\det(\Sigma_{Fm})^{-1/2}\exp\left(-\frac{\phi_{Fm}}{2}(\bx_{Fmr}-\bx_{Fm})\trans\Sigma_{Fm}\inv(\bx_{Fmr}-\bx_{Fm})\right)\times\\
	&\displaystyle\prod_{m=1}^{M}\phi_{Hm}^{\nu_H/2-1}\exp(-\dfrac{\nu_H\phi_{Hm}}{2\phi_H})\displaystyle\prod_{m=1}^{M}\phi_{Fm}^{\nu_F/2-1}\exp(-\dfrac{\nu_F\phi_{Fm}}{2\phi_F}).
	\end{array}
	\]	
	Then, the full conditional distribution of $\blambda$ is
	\[
	\scriptsize
	\begin{array}{rcl}
	\phi_{Hm}\mid\ldots&\sim&Ga\bigg(\dfrac{1}{2}(nR_{Hm}+\nu_H),\dfrac{1}{2}\big\{\displaystyle\sum_{r=1}^{R_{Hm}}(\bx_{Hmr}-\bx_{Hm})\trans\Sigma_{Hm}\inv(\bx_{Hmr}-\bx_{Hm})+\nu_H\phi_H\inv\big\}\bigg),\\
	\phi_{Fm}\mid\ldots&\sim&Ga\bigg(\dfrac{1}{2}(nR_{Fm}+\nu_F),\dfrac{1}{2}\big\{\displaystyle\sum_{r=1}^{R_{Fm}}(\bx_{Fmr}-\bx_{Fm})\trans\Sigma_{Fm}\inv(\bx_{Fmr}-\bx_{Fm})+\nu_F\phi_F\inv\big\}\bigg).\\	
	
	\Pr(\gamma_{Hm}\mid\ldots)&\propto&
	\det(\Sigma_{Hm})^{-R_{Hm}/2}\times\displaystyle\prod_{r=1}^{R_{Hm}} \exp\left(-\frac{\phi_{Hm}}{2}(\bx_{Hmr}-\bx_{Hm})\trans\Sigma_{Hm}\inv(\bx_{Hmr}-\bx_{Hm})\right)\times1_{[0,10^6]}\\
	
	\Pr(\gamma_{Fm}\mid\ldots)&\propto&
	\det(\Sigma_{Fm})^{-R_{Fm}/2}\times\displaystyle\prod_{r=1}^{R_{Fm}} \exp\left(-\frac{\phi_{Fm}}{2}(\bx_{Fmr}-\bx_{Fm})\trans\Sigma_{Fm}\inv(\bx_{Fmr}-\bx_{Fm})\right)\times1_{[0,10^6]}\\
	
	\end{array}
	\]

	The probability of $\btheta$ given $\bW,\bY,\bchi,\bxi,\blambda$ is
	\[
	\scriptsize
	\begin{array}{rl}
	&\Pr(\btheta\mid\bW,\bY,\bchi,\bxi,\blambda)\\
	\propto&\Pr(\bW\mid \bY, \btheta)\Pr(\bY\mid\bxi,\btheta)\Pr(\bchi\mid \btheta,\bxi)\Pr(\blambda\mid\btheta)\Pr(\btheta)\\
	
	\propto
	&\tau^{Nn/2}_W\displaystyle\prod_{i=1}^{N}\exp\left(-\frac{\tau_W}{2}(\by_{Ha}-\bw_i)\trans(\by_{Ha}-\bw_i)\right)\times\\
	&\tau_H^{n/2}\det(\Sigma_H)^{-1/2}\exp\left(-\dfrac{\tau_H}{2\kappa}(\by_H-\bmu_H)\trans\Sigma_H\inv(\by_H-\bmu_H)\right)\times\\
	&\tau_F^{n/2}\det(\Sigma_F)^{-1/2}\exp\left(-\dfrac{\tau_F}{2\kappa}\big\{\by_F-\bmu_F-\beta(\by_H-\bmu_H)\big\}\trans\Sigma_F\inv\big\{\by_F-\bmu_F-\beta(\by_H-\bmu_H)\big\}\right)\times\\
	&\displaystyle\phi^{n/2}_{Ha}\exp\left(-\frac{\phi_{Ha}}{2}(\by_{Ha}-\by_H)\trans(\by_{Ha}-\by_H)\right)\times\\
	&\displaystyle\phi_{Fa}^{n/2}\exp\left(-\frac{\phi_{Fa}}{2}(\by_{Fa}-\by_F)\trans(\by_{Fa}-\by_F)\right)\times\\
	&\tau_H^{Mn/2}\det(V\otimes \Sigma_H)^{-1/2} \exp\left(-\dfrac{\tau_H}{2}  \big\{\vecop(X_{Hm})-\bone_M\otimes\bmu_H\big\}\trans (V\otimes \Sigma_H)\inv \big\{\vecop(X_{Hm})-\bone_M\otimes\bmu_H\big\}\right)\times\\
	&\tau_F^{Mn/2}\det(V\otimes \Sigma_F)^{-1/2} \exp\left(-\dfrac{\tau_F}{2} \big[\vecop(X_{Fm})-\bone_M\otimes\bmu_F-\beta\big\{\vecop(X_{Hm})-\bone_M\otimes\bmu_H\big\}\big]\trans (V\otimes \Sigma_F)\inv \times\right.\\
	&\big[\vecop(X_{Fm})-\bone_M\otimes\bmu_F-\beta\big\{\vecop(X_{Hm})-\bone_M\otimes\bmu_H\big\}\big]\bigg)\times\\
	
	&\displaystyle\prod_{m=1}^{M}\dfrac{\{\nu_H\phi_H\inv/2\}^{\{\nu_H/2\}}}{\Gamma(\nu_H/2)}\phi_{Hm}^{\nu_H/2-1}\exp(-\dfrac{\nu_H\phi_{Hm}}{2\phi_H})\times\displaystyle\prod_{m=1}^{M}\dfrac{\{\nu_F\phi_F\inv/2\}^{\{\nu_F/2\}}}{\Gamma(\nu_F/2)}\phi_{Fm}^{\nu_F/2-1}\exp(-\dfrac{\nu_F\phi_{Fm}}{2\phi_F})\times\\
	&\exp(-\dfrac{10^{-6}}{2}\beta^2)\times\\
	&\tau_W^{(10^{-3}-1)}\exp(-10^{-3}\tau_W)\times\\
	&	\dfrac{\{\nu_H\phi_H\inv/(2\kappa)\}^{\{\nu_H/(2\kappa)\}}}{\Gamma(\nu_H/(2\kappa))}\phi_{Ha}^{\nu_H/(2\kappa)-1}\exp(-\dfrac{\nu_H\phi_{Ha}}{2\kappa\phi_H})\times\\
	&
	\dfrac{\{\nu_F\phi_F\inv/(2\kappa)\}^{\{\nu_F/(2\kappa)\}}}{\Gamma(\nu_F/(2\kappa))}\phi_{Fa}^{\nu_F/(2\kappa)-1}\exp(-\dfrac{\nu_F\phi_{Fa}}{2\kappa\phi_F})\times\\
	&
	\phi_H^{(10^{-3}-1)}\exp(-10^{-3}\phi_H)\times\phi_F^{(10^{-3}-1)}\exp(-10^{-3}\phi_F)\times\\
	&
	\nu_H^{(10^{-3}-1)}\exp(-10^{-3}\nu_H)\times\nu_F^{(10^{-3}-1)}\exp(-10^{-3}\nu_F)\times\\
	&
	\tau_H^{(10^{-3}-1)}\exp(-10^{-3}\tau_H)\times\tau_F^{(10^{-3}-1)}\exp(-10^{-3}\tau_F)\times\\
	&1_{[0,10^6]}(\gamma_H)\times1_{[0,10^6]}(\gamma_F)\times\\
	&\det(V)^{-(M+1+d/2)}\exp\big(-\dfrac{1}{2}\hbox{tr}(d\tilde VV\inv)  \big),\\
	\end{array}
	\]	
	where $1_{[0,10^6]}(\cdot)$ is the indicator function on set $[0,10^6]$ and \texttt{tr} means taking the trace.
	
	Then, the full conditional distribution of $\btheta$ is
	\[
	\scriptsize
	\begin{array}{rcl}
	\tau_W\mid\ldots&\sim&Ga\bigg(\dfrac{Nn}{2}+10^{-3},
	\displaystyle\prod_{i=1}^{N}\frac{1}{2}(\by_{Ha}-\bw_i)\trans(\by_{Ha}-\bw_i)+10^{-3}\bigg),\\
	
	\phi_{Ha}\mid\ldots&\sim&Ga\bigg(
	\dfrac{n}{2}+\dfrac{\nu_H}{2\kappa},\dfrac{1}{2}(\by_{Ha}-\by_H)\trans(\by_{Ha}-\by_H)+\dfrac{\nu_H}{2\kappa\phi_H}\bigg),\\
	
	\phi_{Fa}\mid\ldots&\sim&Ga\bigg(
	\dfrac{n}{2}+\dfrac{\nu_F}{2\kappa},\dfrac{1}{2}(\by_{Fa}-\by_F)\trans(\by_{Fa}-\by_F)+\dfrac{\nu_F}{2\kappa\phi_F}\bigg),\\
	
	\phi_H\mid\ldots&\sim&IG\bigg(10^{-3}+
	\dfrac{\nu_HM}{2}+\dfrac{\nu_H}{2\kappa},10^{-3}+\displaystyle\sum_{m=1}^{M}\dfrac{\nu_H\phi_{Hm}}{2}+\dfrac{\nu_H\phi_{Ha}}{2\kappa}\bigg),\\
	
	\phi_F\mid\ldots&\sim&IG\bigg(10^{-3}+
	\dfrac{\nu_FM}{2}+\dfrac{\nu_F}{2\kappa},10^{-3}+\displaystyle\sum_{m=1}^{M}\dfrac{\nu_F\phi_{Fm}}{2}+\dfrac{\nu_F\phi_{Fa}}{2\kappa}\bigg),\\

	\tau_H\mid\ldots&\sim&Ga\bigg(\dfrac{1}{2}(M+1)n+10^{-3},\\
	&&\dfrac{1}{2\kappa}(\by_H-\bmu_H)\trans\Sigma_H\inv(\by_H-\bmu_H)  
	+\\
	&&\dfrac{1}{2}\hbox{vec}\big(X_{Hm}-\bmu_H\bone_M\trans\big)\trans\hbox{vec}\big(\Sigma\inv_H(X_{Hm}-\bmu_H\bone_M\trans)V\inv\big)+10^{-3}\bigg),\\

	\tau_F\mid\ldots&\sim&Ga\bigg(\dfrac{1}{2}(M+1)n+10^{-3},\\
	&&\dfrac{1}{2\kappa} \big\{\by_F-\bmu_F-\beta(\by_H-\bmu_H)\big\}\trans\Sigma\inv_F\big\{\by_F-\bmu_F-\beta(\by_H-\bmu_H)\big\} 
	+\\
	&&\dfrac{1}{2}\hbox{vec}\big(X_{Fm}-\bmu_F\bone_M\trans-\beta(X_{Hm}-\bmu_H\bone_M\trans)\big)\trans\times\\
	&&\hbox{vec}\big(\Sigma\inv_F\big\{X_{Fm}-\bmu_F\bone_M\trans-\beta(X_{Hm}-\bmu_H\bone_M\trans)\big\}V\inv\big)+10^{-3}\bigg),\\

	V\mid\ldots&\sim&IW\bigg(d\tilde V+\tau_H(X_{Hm}-\bmu_H\bone_M)\trans\Sigma_H\inv(X_{Hm}-\bmu_H\bone_M)\\
	&&+\tau_F\big\{X_{Fm}-\bmu_F\bone_M\trans-\beta(X_{Hm}-\bmu_H\bone_M)\big\}\trans\Sigma_F\inv\big\{X_{Fm}-\bmu_F\bone_M\trans-\beta(X_{Hm}-\bmu_H\bone_M)\big\},\\
	&&2n+M+d+1\bigg),\\
	
	\beta\mid\ldots&\sim&N\big(\dfrac{{v}_{\beta}}{Q_{\beta}},\dfrac{1}{Q_{\beta}}\big)\\
	
	\Pr(\nu_H\mid\ldots)&\propto&
	\displaystyle\prod_{m=1}^{M}\dfrac{\{\nu_H\phi_H\inv/2\}^{\{\nu_H/2\}}}{\Gamma(\nu_H/2)}\phi_{Hm}^{\nu_H/2-1}\exp(-\dfrac{\nu_H\phi_{Hm}}{2\phi_H})\times\\
	&&
	\dfrac{\{\nu_H\phi_H\inv/(2\kappa)\}^{\{\nu_H/(2\kappa)\}}}{\Gamma(\nu_H/(2\kappa))}\phi_{Ha}^{\nu_H/(2\kappa)-1}\exp(-\dfrac{\nu_H\phi_{Ha}}{2\kappa\phi_H})\times\nu_H^{(10^{-3}-1)}\exp(-10^{-3}\nu_H),\\

	\Pr(\nu_F\mid\ldots)&\propto&
	\displaystyle\prod_{m=1}^{M}\dfrac{\{\nu_F\phi_F\inv/2\}^{\{\nu_F/2\}}}{\Gamma(\nu_F/2)}\phi_{Fm}^{\nu_F/2-1}\exp(-\dfrac{\nu_F\phi_{Fm}}{2\phi_F})\times\\
	&&
	\dfrac{\{\nu_F\phi_F\inv/(2\kappa)\}^{\{\nu_F/(2\kappa)\}}}{\Gamma(\nu_F/(2\kappa))}\phi_{Fa}^{\nu_F/(2\kappa)-1}\exp(-\dfrac{\nu_F\phi_{Fa}}{2\kappa\phi_F})\times\nu_F^{(10^{-3}-1)}\exp(-10^{-3}\nu_F),\\
	
	\Pr(\gamma_H\mid\ldots)&\propto&
	\det(\Sigma_H)^{-1/2}\exp\left(-\dfrac{\tau_H}{2\kappa}(\by_H-\bmu_H)\trans\Sigma_H\inv(\by_H-\bmu_H)\right)\times\\
	&&
	\det(\Sigma_H)^{-M/2} \exp\left(-\dfrac{\tau_H}{2}  \big\{\vecop(X_{Hm})-\bone_M\otimes\bmu_H\big\}\trans (V\otimes \Sigma_H)\inv \big\{\vecop(X_{Hm})-\bone_M\otimes\bmu_H\big\}\right)\times1_{[0,10^6]}\\

	\Pr(\gamma_F\mid\ldots)&\propto&
	\det(\Sigma_F)^{-1/2}\exp\left(-\dfrac{\tau_F}{2\kappa}\big\{\by_F-\bmu_F-\beta(\by_H-\bmu_H)\big\}\trans\Sigma_F\inv\big\{\by_F-\bmu_F-\beta(\by_H-\bmu_H)\big\}\right)\times\\
	&&
	\det(\Sigma_F)^{-M/2} \exp\left(-\dfrac{\tau_F}{2} \big[\vecop(X_{Fm})-\bone_M\otimes\bmu_F-\beta\big\{\vecop(X_{Hm})-\bone_M\otimes\bmu_H\big\}\big]\trans (V\otimes \Sigma_F)\inv \times\right.\\
	&&\big[\vecop(X_{Fm})-\bone_M\otimes\bmu_F-\beta\big\{\vecop(X_{Hm})-\bone_M\otimes\bmu_H\big\}\big]\bigg)\times1_{[0,10^6]},\\
	\end{array}
	\]
	where $\nu_H$, $\nu_F$, $\gamma_H$, and $\gamma_F$ do not have a standard from for the full conditional likelihood so a Metropolis-Hasting update is used in the sampling, and
	\[
	\scriptsize
	\begin{array}{rcl}
	{v}_{\beta}&=&\tau_F/(\by_H-\bmu_H)\trans\Sigma_F\inv(\by_F-\bmu_F)/\kappa+\\
	&&\tau_F\hbox{vec}(X_{Hm}-\bmu_H\bone\trans_M)\trans\hbox{vec}\{\Sigma\inv_F(X_{Fm}-\bmu_F\bone_M\trans)V\inv\}\\
	Q_{\beta}&=&\tau_F(\by_H-\bmu_H)\trans\Sigma_F\inv(\by_H-\bmu_H)/\kappa\\
	&&+\tau_F\hbox{vec}\big(X_{Hm}-\bmu_H\bone_M\trans\big)\trans\hbox{vec}\big(\Sigma_F\inv(X_{Hm}-\bmu_H\bone_M\trans)V\inv\big) +10^{-6}\\
	\end{array}
	\]
	It is also noteworthy that we scale the matrix $V$ so that $V_{(1,1)}$ equals one in each step of the updates to make $\tau_H$ and $\tau_F$ identifiable.

\bibliography{ref}
\bibliographystyle{chicago}

\end{document}